\documentclass[manuscript]{aastex}

\usepackage{txfonts}
\usepackage{longtable}
\usepackage{rotating}
\usepackage{natbib}
\usepackage{graphicx}
\usepackage{graphics}
\usepackage{psfrag}
\usepackage{amssymb}
\bibpunct{(}{)}{;}{a}{}{,}
\def\cs{count~s$^{-1}$}
\def\Teff{$T_{\mathrm{eff}}$}

\def\kms{$\mathrm{km\,s}^{-1}$}

\def\llm{{\sc LLmodels}}

\shorttitle{Near-UV Absorption, Chromospheric Activity, and Star-Planet Interactions in WASP-12.}
\shortauthors{Haswell et al.}

\begin{document}

\title{Near-UV Absorption, Chromospheric Activity, and Star-Planet Interactions
in the WASP-12 system.\altaffilmark{1}}
\altaffiltext{1}{Based on observations made with the NASA/ESA Hubble Space
Telescope, obtained from MAST at the Space Telescope Science Institute, which is
operated by the Association of Universities for Research in Astronomy, Inc.,
under NASA contract NAS 5-26555. These observations are associated with
programs \#11651 and \#11673.}

\author{C.A. Haswell and L. Fossati}
\affil{Department of Physical Sciences, The Open University,
	Walton Hall, Milton Keynes MK7 6AA, UK}
\email{C.A.Haswell@open.ac.uk,l.fossati@open.ac.uk}
\author{T. Ayres, K. France  and C. S. Froning}
\affil{Center for Astrophysics and Space Astronomy,
	University of Colorado, 593 UCB, Boulder, CO 80309-0593}
\email{cynthia.froning@colorado.edu}
\author{S. Holmes, U.C. Kolb, R. Busuttil}
\affil{Department of Physical Sciences, The Open University,
	Walton Hall, Milton Keynes MK7 6AA, UK}
\author{R.A. Street}
\affil{Las Cumbres Observatory}
\author{L. Hebb}
\affil{Department of Physics and Astronomy, Vanderbilt University, 6301
	Stevenson Center Nashville, TN 37235, USA}
\email{leslie.hebb@vanderbilt.edu}
\author{A. Collier Cameron, B. Enoch}
\affil{School of Physics and Astronomy, University of St Andrews,
North Haugh, St Andrews, Fife, Scotland, KY16 9SS
}
\author{V. Burwitz}
\affil{Max Planck Institut f{\"u}r extraterrestrische Physik, Giessenbachstrasse, 85748 Garching, Germany}
\author{J. Rodriguez}
\affil{Observatori Astron\'omic de Mallorca, Cam\'i de l'Observatori, 07144 Costitx, Mallorca, Spain}
\author{R.G. West}
\affil{Department of Physics \& Astronomy, University of Leicester, Leicester
LE1 7RH, UK}
\author{D. Pollacco, P.J. Wheatley}
\affil{Department of Physics, University of Warwick, Coventry CV4 7AL, UK}
\author{A. Carter}
\affil{Department of Physical Sciences, The Open University,
	Walton Hall, Milton Keynes MK7 6AA, UK}

\begin{abstract}
Extended gas clouds have been previously detected surrounding the brightest known close-in hot
Jupiter exoplanets, HD\,209458\,b and HD\,189733\,b;
we observed the distant but more extreme close-in hot Jupiter system, WASP-12,
with HST.
Near-UV (NUV) transits up to three times deeper
than the optical transit of WASP-12\,b reveal
extensive diffuse gas, extending well beyond the Roche lobe.
The distribution
of absorbing gas varies between visits.
The deepest NUV transits are at
wavelength ranges with strong photospheric absorption, implying the
absorbing gas may have temperature and composition similar to the stellar photosphere.
Our spectra reveal significantly enhanced absorption (greater than $ 3 \sigma$ below
the median) at $\sim 200$  individual wavelengths on each of two HST visits;
65 of these wavelengths are consistent between the two visits, using a strict criterion for velocity matching which
excludes matches with velocity shifts exceeding $\sim 20\,{\rm km\,s^{-1}}$.
Excess transit depths are robustly detected throughout the inner wings of the \ion{Mg}{2} resonance lines
independently on both HST visits.
We detected absorption in \ion{Fe}{2}
2586\AA ,
the heaviest species yet detected in an exoplanet transit.
The  \ion{Mg}{2} line cores have zero flux, emission cores exhibited by
every other
observed star of similar age and spectral type are conspicuously absent.
WASP-12
probably produces normal \ion{Mg}{2} profiles, but the inner portions of these strong
resonance lines are likely affected by extrinsic absorption.  The required Mg$^{+}$ column is an order of magnitude greater than
expected from the ISM,
though we cannot completely dismiss that possibility.  A more plausible
source of absorption is gas lost by WASP-12\,b.
We show that planetary mass loss
can produce the required column.
Our Visit 2 NUV light curves show evidence for a stellar flare.
We show that some of the possible transit detections in resonance lines of rare elements may be
due instead to non-resonant transitions in common species.
We present optical observations and update the transit ephemeris.
\end{abstract}
%


%\keywords{globular clusters: general --- globular clusters: individual
%(NGC 6397, NGC 6624, NGC 7078, Terzan 8}

\section{Introduction}\label{intro}
WASP-12\,b is one of the most extreme of the hot Jupiter exoplanets: it orbits only one stellar diameter away from the photosphere of its late F type host star, and is thus one of the most irradiated planets known \citep{hebb2009}.
The brightest hot Jupiters, HD 189733\,b and HD 209458\,b, are known to be surrounded by extended clouds of absorbing gas which have been
detected through observations of their transits in the far UV \citep[FUV;][]{vidal03,vidal04,benjaffel07,vidal08,lecav10,linsky10,2012...26A...543L...4L}.
These FUV transits occult the patchy and time-variable chromospheric emission from the star \citep{Haswell10}: the FUV continuum from even the nearest transiting-planet host stars is faint \citep[see e.g.][]{linsky10} . The FUV transit light curve therefore depends on the locus the planet follows across this constantly changing irregular chromospheric emission distribution and also depends on the velocity range of the emission line adopted for analysis \citep{benjaffel07,vidal08}.

Several mechanisms have been proposed to explain the extended absorbing gas.
A
hydrodynamic `blow-off' of the planet's outer atmosphere caused by the
intense irradiation was the first and most obvious \citep{vidal04}.
This model has been extensively discussed and refined, most recently by \citet{guo11} and \citet{Ehrenreich11}.
\citet{adams11} considers how these outflows might be influenced by the planetary
magnetic field, while \citet{lecav04,lecav07} and \citet{erkaev07} consider the effect of the Roche equipotentials
on the outflow.

\citet{li10} suggested that for eccentric hot Jupiters,  tidal disruption might drive mass loss orders of magnitude
 greater than that from irradiation-driven hydrodynamic outflow. This was motivated by the implications in the
discovery paper
that WASP-12\,b had
a significantly non-zero eccentricity \citep{hebb2009}.
The most recent measurements of the orbital eccentricity of WASP-12\,b
are, however, consistent with zero eccentricity \citep{lm09,husnoo11}, so tidal disruption is
not expected to play a significant role for WASP-12\,b. It could nonetheless play an important role
for hot Jupiters found to have  non-zero eccentricity, for example due to perturbations by
other planets in the system.

 A third
explanation proposes that the planet is surrounded by a cloud of energetic
neutral atoms caused by interactions with the host star's stellar wind
\citep{ena08,ena10}. \citet{2012arXiv1206.5003T} simulate these colliding
winds reproducing the observed Lyman$\alpha$ transit profiles. \citet{Raiput11} report
the formation of energetic neutral atoms from ion-impact dissociation
of water in the laboratory, and suggest this process may contribute to the
absorbing gas in hot Jupiter exospheres.

\citet{LHH10} considered mass loss via Roche lobe overflow in WASP-12, while
others \citep[e.g.][]{vidotto10} suggest
the absorbing gas may simply be
entrained material from the stellar corona,
in which case it provides an opportunity to measure the planetary magnetic field.
\citet{llama11} model the radiative transfer through compressed material behind
 the magnetospheric bow shock produced in this scenario.

Since WASP-12\,b is such an extreme hot Jupiter it was
an
obvious target for observations to examine the properties of the extended absorbing gas
around these systems. The UV spectral region contains many resonance lines
of common elements, including Lyman $\alpha$ which provided the first detection of this
phenomenon \citep{vidal03}. These resonance lines provide a sensitive probe for the presence of
absorbing gas. WASP-12's distance is 300\,pc or more \citep{fossati2010b},
and this means the predicted FUV emission would be extremely hard to detect.
 Thus  we chose to observe in the near UV (NUV)
using the COS spectrograph on  HST to cover the strong \ion{Mg}{2} resonance lines
and a host of other spectral lines from a rich variety of chemical elements.
Our choice of the NUV spectral region has the additional benefit that our
transit light curves are measured using the well-understood intensity
distribution from the stellar photosphere, which is expected to remain constant
with time. In the NUV we are also able to detect absorption from gas with
any velocity: in the FUV, gas ceases to produce an observable effect once its velocity
exceeds the velocity width of the stellar chromospheric line emission  which is
being absorbed.

Our observations comprised two visits,
and we published a short paper on the results from the first visit in \citet[][hereafter Paper 1]{paper1}.
In Paper 1 we presented light curves revealing an exosphere which appears to overfill its Roche lobe
and wavelength resolved data, identifying
enhanced transit depths at the wavelengths of
a number of resonance lines.

In this paper, we report on our full complement of HST/COS NUV observations of WASP-12 and on
contemporaneous ground-based optical photometry. In \S\ref{obs} we describe our observations
and data reduction. \S\ref{sec:res} gives our observational results and develops their
interpretation. \S\ref{sec:conc} critically examines the interpretation suggested by our observational
results and presents calculations demonstrating their plausibility.
While the conclusions of Paper 1 are not significantly changed by
our re-examination with the benefit of more extensive data, the new data demands a
more complex interpretation than our first analysis in Paper 1 suggested.

\section{Observations and data reduction}\label{obs}
We observed WASP-12 with HST/ COS;
and from the ground with PIRATE \citep{Holmes2011}, Faulkes Telescope North, and the James Gregory Telescope. These observations are described in \S\S\ref{hstobs},
\ref{sec:pirateobs}, \ref{sec:FTNobs} and  ~\ref{sec:JGTobs} respectively.

\subsection{HST/COS observations}
\label{hstobs}

%\clearpage
\begin{table}
%\begin{center}
\caption{\label{table_hstobs} The temporal and wavelength coverage of our
two HST/COS Visits.}
%\label{table_bands}
\begin{tabular}{l|c|c}
\tableline\tableline
Visit & Wavelength   & Temporal  coverage      \\
      & range [\AA] & [HJD-2455000] \\
\tableline
  & NUVA: 2539--2580    & 099.274652 - 099.296111 (orb1) \\
1 & NUVB: 2655--2696    & 099.330823 - 099.369516 (orb2) \\
  & NUVC: 2770--2811    & 099.397468 - 099.432689 (orb3) \\
  &                     & 099.464090 - 099.499311 (orb4) \\
  &                     & 099.530735 - 099.565956 (orb5) \\
\tableline
  & NUVA: 2551--2594    & 283.695216 - 283.722230 (orb1) \\
2 & NUVB: 2669--2711    & 283.753387 - 283.788606 (orb2) \\
  & NUVC: 2789--2829    & 283.819948 - 283.855168 (orb3) \\
  &                     & 283.886510 - 283.921729 (orb4) \\
  &                     & 283.953072 - 283.988291 (orb5) \\
%2 & NUVA:     & $m_{b}$ & 2782.75 - 2787.5  & $w_{b}$ & 2782.75 - 2783.75 \\
\tableline
\end{tabular}
\end{table}

COS
 is a slitless spectrograph which we used to obtain R$\sim$20\,000
 NUV spectroscopy.
COS maximises the spectroscopic sensitivity of HST in the UV; a full instrument description and on-orbit performance characteristics can be
found in \citet{osterman} and \citet{2012ApJ...744...60G}.

Our HST observations are summarised in Table~\ref{table_hstobs}.  Visits 1 and 2 each comprised
five consecutive HST orbits and were executed on 2009 September 24/25 and 2010 March 28 respectively.
A first analysis and brief discussion of Visit\,1 was given in Paper 1.
For Visit\,1 we used the NUV G285M grating at the 2676\,\AA\ setting,
which provides non-contiguous spectra over three wavelength ranges; see Table~\ref{table_hstobs}.
For Visit 2 we used the 2695\,\AA\ setting of the same grating which gives a slight shift to the red
as detailed in Table~\ref{table_hstobs}.
Our set-up for the two visits was otherwise identical, with
spectral resolution of $R\sim$20\,000,
without the use of offset FP-POS positions.
We obtained simultaneous lamp spectra during science exposures using ``FLASH = YES''.

The settings for Visit\,1 optimized coverage of the
region around the core of the strong \ion{Mg}{2} UV resonance lines; for Visit\,2 we
chose to cover the \ion{Fe}{2} resonance line at $\lambda $2586\,\AA\
% ref Kroll & Koch 1987, A&ASupp, 67, 225
while
maximising the overlap in wavelength between the two
sets of observations. Obtaining a homogeneous data-set may have
had advantages, as we are photon starved, and the behavior of the WASP-12 system
appears different on the two visits. The wavelength shift means that only
a subset of our spectral coverage can be combined to make
light curves which can be straightforwardly interpreted.
In \S\S\ref{sec:abs} and \ref{sec:flare}, however, we use the \ion{Fe}{2} resonance line
to disentangle a complicated story, and it is decisive in one of the major interpretative
questions of this paper.

In each visit the exposure time was 2334\,s in the first HST orbit and about 3000\,s  per subsequent
HST orbit. We designed the timings of the two visits so they interleave to give full phase coverage of the transit with out of transit (OOT)
coverage before ingress and after egress.
 Our observations
were obtained in TIME-TAG observing mode, in which individual photon events
are recorded at a resolution of 32\,ms, but our effective time-resolution is much poorer
due to the faintness of the target.
The count rates summed over wavelength are $\sim 10$\,\cs; $\sim 28$\,\cs; and $\sim 13$\,\cs \,
 respectively for the NUVA, NUVB, and NUVC ranges (see Table~\ref{table_hstobs} for definitions).
The data were reduced using the calibration files provided by STScI; these were updated following Paper 1, so we re-reduced all the HST data using the updated calibration files for the present paper.
We downloaded data from MAST\footnote{\tt http://archive.stsci.edu/} adopting
CALCOS V.2.11b\footnote{See the COS Data Handbook for more information on
CALCOS. The version of CALCOS we used did not correct for declining sensitivity, while more
recent versions now do.
%: \tt http://www.stsci.edu/hst/cos/documents/handbooks/\\datahandbook/COS\_longdhbcover.html.
}
for calibration.
The COS data comprise exposures of the science target with corresponding
lamp spectra displaced in the spatial direction. The effects
of the new reduction on the data presented in Paper 1 are shown at the end
of this section.

The reduced Visit\,2 data as downloaded from the MAST archive suffered from a
problem in the extraction of the science spectra  and in the determination
of the background level. We characterised the location of the
science spectra by fitting Gaussians to the profile in the spatial direction.
The center of the extraction box for the NUVA
region, as defined in the extraction reference file, is about
5 pixels from the central peak of the science spectrum.
Similarly, the center of one
of the two extraction boxes used by CALCOS to compute the background level
is misplaced so that a portion of the science spectrum falls into the
background extraction box.
To solve this problem, we compared the position of the central pixel of each of the NUVA, NUVB and NUVC
spectra as given
in the reference file for different settings of the same grating. The prescribed
central
pixel differs for different settings, but NUVA, NUVB and NUVC are always at the same {\it relative}
positions, except for the
2695\,\AA\ setting. For this reason we simply corrected the extraction box for
the NUVA and background region by adopting the same shift as for the
NUVB and NUVC regions, we then re-ran CALCOS and the extraction succeeded.

%--------------------------------------------------------------------
\begin{figure}
\begin{center}
\includegraphics[width=100mm]{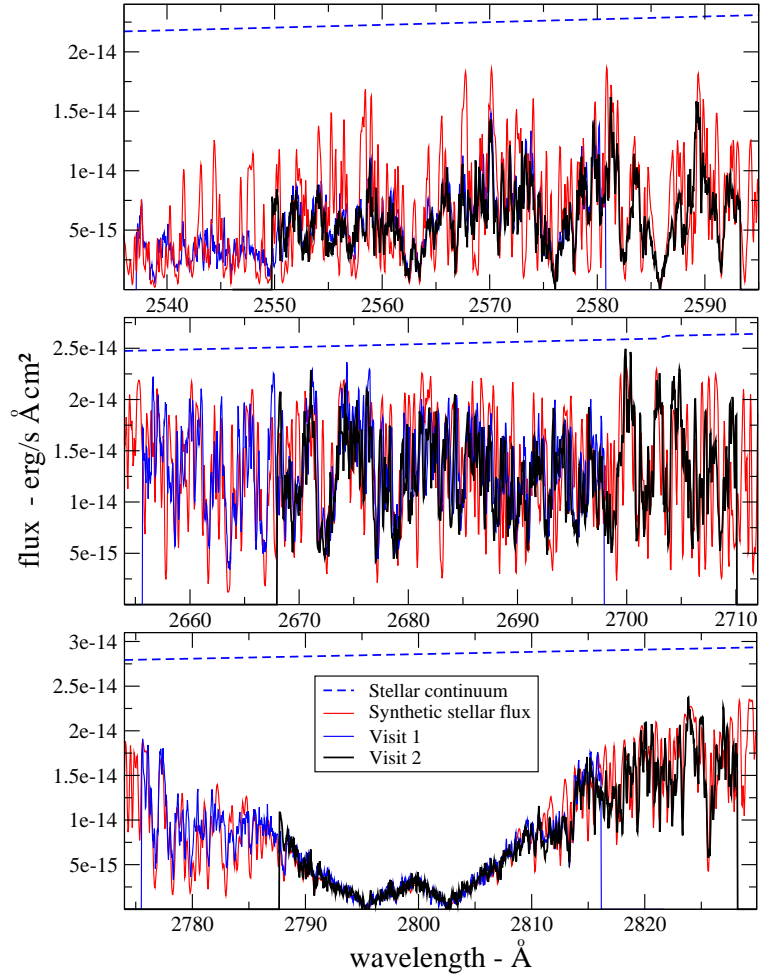}
\caption{HST/COS NUV spectra of WASP-12 (Visit 1: blue line;
Visit 2: black
line) and
a scaled model WASP-12 stellar spectrum (red line).
The stellar continuum (dashed blue line) lies well above the observed flux due to  overlapping
stellar absorption features throughout the spectrum. The mean flux is slightly lower in Visit 2;
 see text. The three panels show the NUVA, NUVB, and NUVC spectral regions.
}
\label{flux-v1ev2}
\end{center}
\end{figure}
%--------------------------------------------------------------------

Figure~\ref{flux-v1ev2} shows the flux-calibrated grand sum spectra obtained in
Visit 1 and Visit 2 in comparison with synthetic fluxes, as discussed
in \S\ref{nuvspec}. The mean fluxes are $9 \pm 1$\%, $7 \pm 1$\%
and $3.5 \pm 1.0$\% lower in Visit\,2 for NUVA, NUVB, and NUVC respectively.
This is largely attributable to the declining throughput \citep{Osten}, but also due to the varying mean obscuration
of the WASP-12 stellar flux by the planet and diffuse gas. We will discuss
a likely third contributor to varying fluxes in \S\ref{sec:abs}.

In our time-series analysis we used the {\it count rates} obtained after
background subtraction, rather than the flux calibrated spectra. The high
quality flat-field and the relatively low background of the NUV channel
mean the uncertainties are dominated by Poisson statistics. The resulting
signal to noise ratio (SNR) in our extracted NUVB spectra is $\sim$8 per pixel for
each 3000\,sec exposure. This is consistent with the expected photon counting noise,
and propagates to a fractional uncertainty of $\sim 0.007$ when a 3000\,s
exposure is averaged.

%--------------------------------------------------------------------
\begin{figure}
\begin{center}
\includegraphics[width=100mm]{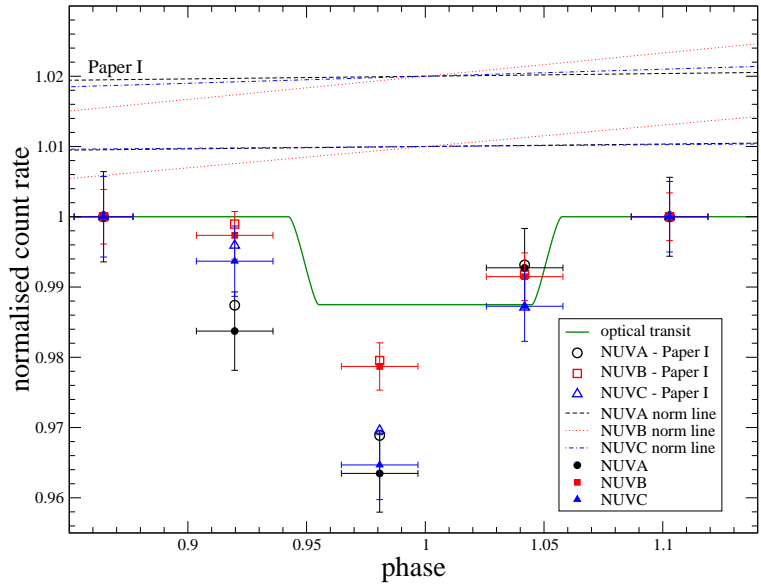}
\caption{The Visit 1 results from Paper 1 (open symbols) compared to the same data
re-reduced with the updated calibration files (closed symbols).
The normalisation lines indicate the tilt which has been removed from the
data in setting the first and last point to unity (i.e. the first two points in the NUVB
light curve were moved slightly upwards).
The upper set of
normalisation lines correspond to the Paper 1 reduction. The green line shows the model
transit of the optically opaque planet across a uniform brightness stellar disc.}
\label{lc_norm-v1}
\end{center}
\end{figure}
%--------------------------------------------------------------------
Figure~\ref{lc_norm-v1} shows our Visit 1 light curves. These were produced by summing across wavelength
for each of NUVA, NUVB and NUVC to produce a photometric point
for each HST orbit.
Fig.~\ref{lc_norm-v1} compares
the results from Paper 1 with those from our revised calibration.
The first and last
data points for both reductions are set to unity with the  normalisation procedure applied  in Paper 1;
all the remaining points are within 1$\sigma$ of their previous positions.
For all three wavelength regions, the normalisation lines (see Paper 1)
are slightly improved (flatter) with the new calibration files.
The normalisation applied does not significantly affect the shape of the observed NUV
transit, as no point moves by as much as 1$\sigma$.

The NUVA point immediately preceding optical ingress and the NUVA and NUVC points
at mid-transit have been moved down almost $ 1 \sigma$ relative to the reduction
presented in Paper 1. The refined reduction therefore strengthens the case (made in Paper 1)
for an early ingress and an extended asymmetric cloud of absorbing gas: the NUVA and NUVC
transit depths are now $\sim 4 \sigma$ below the depth of the optical transit. When we consider the data
from
both visits together, however, these conclusions become more complicated, as discussed in \S\ref{sec:nuv_lc}.

\subsection{PIRATE optical photometry}
\label{sec:pirateobs}
%\clearpage
\begin{table}
%\begin{center}
\caption{\label{table_pirateobs} WASP-12 observations with PIRATE.}
%\label{table_bands}
\begin{tabular}{llllr}
\tableline\tableline
Date &               HJD start  & HJD end & Exp. & Comments    \\
(BON)&       -2455000                   & -2455000        & Time (s)     &              \\
\tableline
15/10/2009  &   121.53078   &  121.68142  & 120s  &   OOT only\\
18/10/2009  &   123.49117   &  123.67301  & 120s  &   OOT only\\
11/11/2009  &   147.47507   &  147.62549  & 120s  &   Egress \& OOT. \\
            &                   &                 &       &   High humidity ($>$ 90\%)\\
12/11/2009  &   148.47187   &  148.51421  &  60s   &   OOT \& ingress\\
22/11/2009  &   158.37221   &  158.69467  &  60s   &   Egress \& OOT\\
23/11/2009  &   159.44126   &  159.59863  &  60s   &   Mid-transit, egress \& OOT\\
13/01/2011  &   575.26270   &  575.36230  &  45s   &   Used PIRATE Mk. 2 \\
            &                   &                 &        &   \citet{Holmes2011}\\
\tableline
\end{tabular}
\end{table}

 %\label{table_pirateobs}
In conjunction with our first HST visit we undertook a program of optical
photometry with The Open University's remotely operable optical telescope, PIRATE \citep{Holmes2011}.
The goal of these observations was to verify or, if necessary,
update the WASP-12\,b orbital ephemeris. The PIRATE observations are logged in Table~\ref{table_pirateobs}.
We used the R filter on all 7 nights, and either
120s, 60s, or 45s exposures. The dead time between exposures is 15s, resulting
in a cadence of 135s, 75s, or 60s. We captured one ingress, four egresses, and on two nights OOT coverage only. The control computer's real-time clock was synchronized every hour with a USNO time server, typically requiring 11-13s correction each time.
Light curve precision was  2.5-2.8 mmag, which we achieved through an
iterative comparison-ensemble procedure \citep{Holmes2011}. The light curve of one egress is published
in \citet[][their Fig. 7]{Holmes2011}, %(their Fig. 7),
and we present our phase-folded optical light curve in
Sect.~\ref{sec:eph} below.
%\citep[][hereafter Paper 1]{fossati2010}

\subsection{Faulkes Telescope North optical photometry}
\label{sec:FTNobs}
The Spectral camera on the 2m Faulkes Telescope North, Hawai'i, was
used to observe WASP-12 on 2010 December 17 UT through a Bessell-B
filter.  The Spectral camera is a 4096x4096 pixel Fairchild CCD486 BI
which was used in bin 2x2 mode giving a 0.304 arcsec/pixel scale over the
$10\times 10$ arcmin field of view.
We used a 22\,s exposure time, which gave a cadence of about 47\,s.
The start and end time of the observations were HJD\_UTC 2455547.913994473
and
2455548.085096728 respectively.
The timeseries data were preprocessed
using the standard ARI pipeline and aperture photometry was performed
using IRAF/DAOphot.

\subsection{James Gregory Telescope photometry}
\label{sec:JGTobs}
A partial transit was obtained in photometric conditions on 30 November 2008 with a 1k $\times$ 1k CCD detector mounted on the James Gregory Telescope (JGT), a $0.95$\,m Schmidt-Cassegrain telescope at the University of St Andrews.
The full field of view (FOV) of the detector is $17^{\prime}$ with a 1 arcsec/pixel platescale, however vignetting and aberrations create an effective FOV for photometry of approximately $10^{\prime}$.
Cousins-R band exposures of 90s duration were taken over 3 hours from  HJD\_UTC = 2454801.388296 to 2454801.517059, with a total of 104 photometric points. Standard image processing,
source detection and aperture photometry were performed using the Cambridge Astronomical Survey Unit catalogue extraction software \citep{irwinlewis01}. The software has been compared with SExtractor  and found to be very similar in the completeness, astrometry and photometry tests.   We adopted an 8 pixel radius aperture to match the  typical seeing for the night (FWHM $\sim 8^{\prime\prime}$).    Differential photometry was generated from the instrumental magnitudes using the combined flux of 7 nearby stars of similar brightness to WASP-12.

\section{Results and Interpretation}
\label{sec:res}
\subsection{The Orbital Ephemeris}
\label{sec:eph}
The elapsed time between the WASP-12\,b discovery paper photometry \citep{hebb2009} and our two HST visits was 18 months and two years respectively. Since we anticipated the interpretation of our NUV time-series data could be critically dependent on the WASP-12 orbital ephemeris, we re-examined it.
We took our PIRATE, FTN and JGT data, the
Liverpool Telescope Z band from \citet{hebb2009},
together with all available SuperWASP data, and determined a new linear ephemeris using the MCMC fitting routine of \citet{acc2007}, updated by \citet{enoch2010}, as used by the SuperWASP consortium. We obtained a new ephemeris of
\begin{equation}\label{eq:eph}
T_{\rm mid}({\rm HJD}) = 2454852.7739 ^{+0.00014}_{-0.00014} + N \times 1.09142206^{+0.00000033}_{-0.00000031}
\end{equation}
Excluding the SuperWASP data and those observations that did not capture any part of the transit, we then examined the data for transit timing variations (TTVs).
We phase-folded the 15 remaining light curves, and renormalised each observation against a model
light-curve appropriate to its color using $\chi^2$ minimisation.
A phase offset from Eqn.~\ref{eq:eph}
was determined by $\chi^2$ minimisation for each individual light curve.
These offsets define the observed (`O') mid-transit time for each observation
and the deviations (`O-C') from the mid-transit time calculated from Eqn.~\ref{eq:eph} (`C') are tabulated
in Table~\ref{tab:omc}.

\begin{table}
\caption{\label{tab:omc} Transit mid-times}
\begin{tabular}{lllr}
\tableline
\tableline
$ T_{\rm mid}$ (`O')&  $ \sigma (T_{\rm mid})$ & `O-C'  & Source \\
    (HJD- 2450000)       &       (s)             &   (s)  &        \\
\tableline

4515.52455 & 11.32 & 5.65793 & LT (Z) \\
4801.47593 & 35.83 & -98.07082 & JGT (R)\\
5123.44587 & 60.35 & -60.35127 & PIRATE (R)\\
5147.45781 & 77.33 & -3.77195 & PIRATE (R)\\
5148.55083 & 113.16 & 133.90439 & PIRATE (R)\\
5158.37107 & 49.04 & -86.75496 & PIRATE (R)\\
5159.46575 & 52.81 & 194.25566 & PIRATE (R) \\
5548.00980 & 20.75 & 3.77195 & FTN (B)\\
5575.29570 & 54.69 & 33.94759 & PIRATE (R) \\

\tableline

\end{tabular}
\end{table}

 %this is 27 june vintage
Our linear ephemeris is well constrained at either end by precise measurements from  Liverpool Telescope and the Faulkes Telescope North.
\citet{Mac11} observed two transits and suggested their measurements might constitute TTV detections.
We note that our new ephemeris is consistent with the \citet{Chan11} transit timings and
 that
we suspect the error estimates of \citet{Mac11} are rather optimistic.
Other light curves gathered with comparable facilities lead to larger timing uncertainties.

None of the PIRATE or JGT observations  captured complete transits, which is one reason why they have large timing uncertainties.
We would not claim a detection of TTVs/TDVs based on these data alone. WASP-12\,b may exhibit TTVs with amplitudes of
$\sim 200$\,s. Further high quality light curves are needed to address this hypothesis.

%--------------------------------------------------------------------
\begin{figure}
\begin{center}
\includegraphics[width=\hsize,clip]{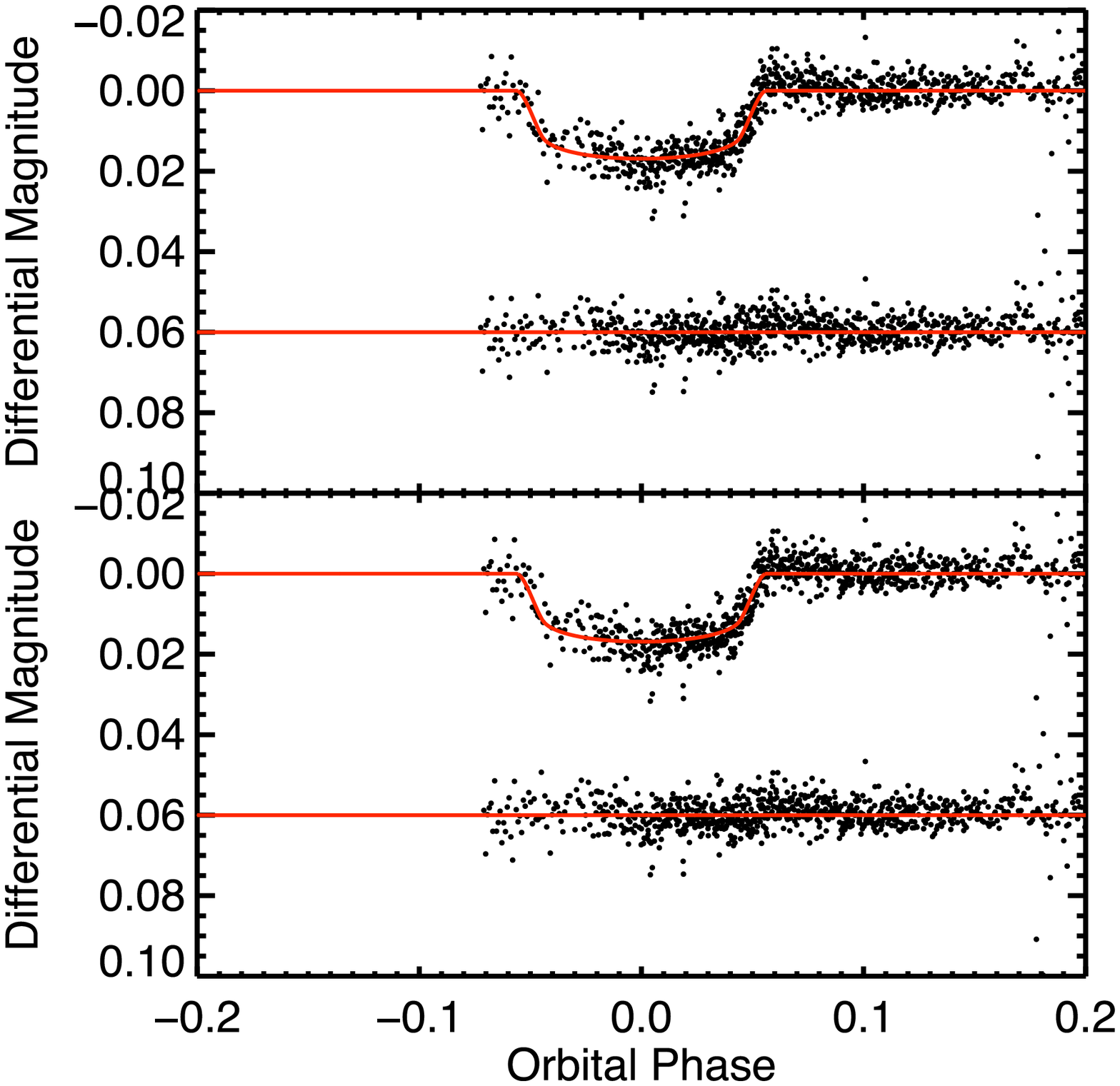}
\caption{PIRATE R band data folded about our linear ephemeris (Eqn.~\ref{eq:eph}); the residuals after subtracting the
R band model from the SuperWASP MCMC run (red line) are shown offset.
 Top panel: with phase-offsets corresponding to the O-C values applied to the data. Bottom panel: without the phase-offsets applied.
}
\label{fig:opt_fold}
\end{center}
\end{figure}
%--------------------------------------------------------------------

\subsection{NUV Spectroscopic coverage}
\label{nuvspec}

Figure~\ref{flux-v1ev2} shows the grand total flux calibrated spectra for our two HST visits. The Visit 2
spectral coverage is redwards of the Visit 1 coverage, and the two ranges overlap
for about half of the total coverage. As noted in \S\ref{hstobs} the Visit 2 spectra
have slightly lower fluxes, largely due to the (uncalibrated) decline in COS throughput.
Both spectra are RV-corrected to the WASP-12 rest frame by line profile fitting to the
two spectra, thus determining an empirical wavelength offset between the
two visits.
This correction was applied before generating Figure~\ref{flux-v1ev2} and subsequent figures.

Figure~\ref{flux-v1ev2}
also shows the stellar continuum level (dashed line).
There is prodigious stellar photospheric absorption in WASP-12's atmosphere throughout; no point
in our NUV spectra is consistent with unabsorbed continuum.
Figure~\ref{flux-v1ev2} shows synthetic fluxes calculated using the
\llm\ stellar model atmosphere code \citep{llm},  assuming the fundamental parameters,
metallicity and detailed abundance pattern given by \citet{fossati2010b}. \llm\ assumes Local
Thermodynamical Equilibrium (LTE) for all calculations, adopts a
plane-parallel geometry, and uses direct sampling of the line opacity
allowing the computation of model atmospheres with individualised (not scaled
to solar) abundance patterns. We used the VALD database
\citep{vald1,vald2,vald3} for the atomic line parameters.  The resolution applied to the synthetic fluxes matches that of
COS, but we did not model the peculiar line spread
function (LSF)
of HST\footnote{\tt
http://www.stsci.edu/hst/cos/documents/newsletters/full\_stories/\\2010\_02/1002\_cos\_lsf.html}.
The differences between the synthetic
and observed spectra are predominantly caused by the HST LSF.

All three regions are affected  throughout by many blended photospheric
absorption lines. We observe no unabsorbed stellar continuum.
The NUVA region is strongly absorbed by many overlapping lines, with lines of MgI and FeI predominating;
the NUVB
region for both visits is closest to the continuum;
while the NUVC region is dominated by  absorption in the broad wings of the strong \ion{Mg}{2} doublet at 2795.5\,\AA\ and 2802.7\,\AA.

\subsection{NUV Transit Light Curve}
\label{sec:nuv_lc}
%\clearpage
\begin{table}
%\begin{center}
\caption{\label{table_chi} Simple transit model fits to NUV light curves. The free parameters are the
phase offset from the optical ephemeris, $\Delta \phi$, and the
radius, $R$, of a circular occulting disc, expressed as a fraction of the stellar radius. We constrained
our search to $R/R_{*}\ge 0.1119 = R_{\rm P}/R_{*}$.}
%\label{table_bands}
\begin{tabular}{cccccc}
\tableline\tableline
BAND  & VISIT(S)  & $\chi ^{2}_{\nu}$ & $N_{params} -$ & $R/R_{*}$ & $\Delta \phi$    \\
      &           &                   & $N_{data} $    &           &  \\
\tableline
A    &   1       &  0.276           &2 & 0.189     &   -0.0279   \\ 
B    &   1       &  0.864           & 2& 0.146    &   -0.0186   \\  
C    &   1       &  0.276           & 2& 0.186     &   -0.0170   \\ 
A    &   2       &  7.77            & 2& 0.118    &    -0.124   \\ 
B    &   2       &  4.11            &2 & 0.113     &    0.0050   \\ 
C    &   2       &  8.32            & 2& 0.136     &    -0.1271 \\
A    &   2NF     &  11.5            &1 & 0.237     &    -0.0475   \\ 
B    &   2NF     &  8.22            & 1& 0.116     &    -0.0204   \\ 
C    &   2NF     &  15.8            &1 & 0.224     &    -0.0499   \\ 
AO    &   1,2     &   4.002         & 6& 0.147     &    -0.0241   \\ 
BO    &   1,2     &   1.823         &6 & 0.129     &    0.0034    \\ 
CO    &   1,2     &   4.74          &6 & 0.122     &    0.0198    \\ 
AO   &  1,2NF    &   4.38           &5 & 0.1633    &    -0.0268    \\ 
BO   &  1,2NF    &   1.25           & 5&         0.1415    &    -0.0159     \\ 
CO   &  1,2NF    &   3.53           & 5&    0.1424   &    -0.0368     \\ 
\tableline
\end{tabular}
\end{table}

\begin{figure}
\begin{center}
\includegraphics[width=70mm]{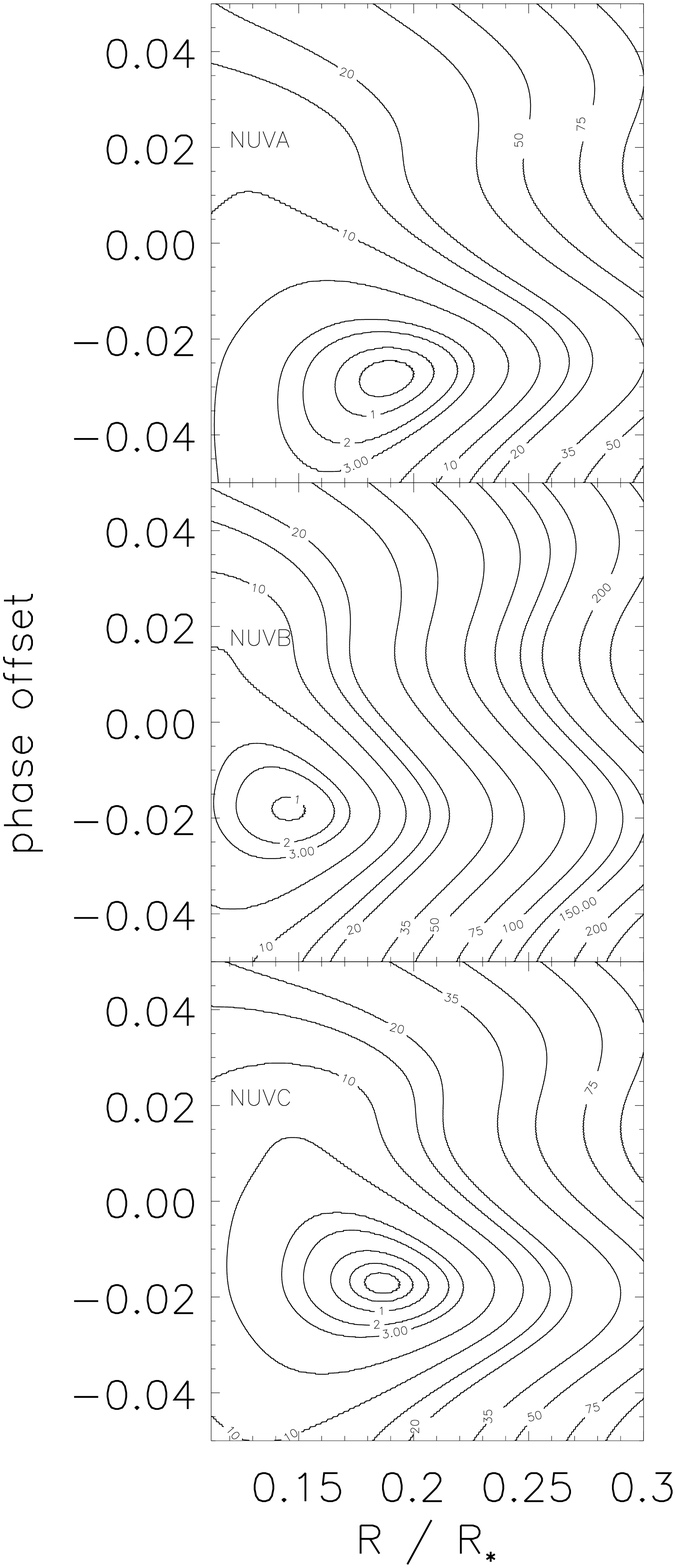}
\caption{$\chi_{\nu} ^2$ grids from fitting model
transits to our Visit 1 NUV data.
We model circular opaque discs
transiting a uniform brightness stellar disc. By allowing
the normalisation of the model light curves to vary
we find the best-fit transit for each value of phase and radius ratio without making any assumptions
about which of the empirical points represent out of transit coverage.
All three light curves are best fit by transits which occur early, and the
best fit models in all cases correspond to discs much larger than that
of the optical planet: $R_{\rm P} / R_{*} = 0.1119$, which is the smallest value of
$R / R_{*}$ shown.
}
\label{fig:chi_nuv}
\end{center}
\end{figure}

\begin{figure}
\begin{center}
\includegraphics[width=17cm]{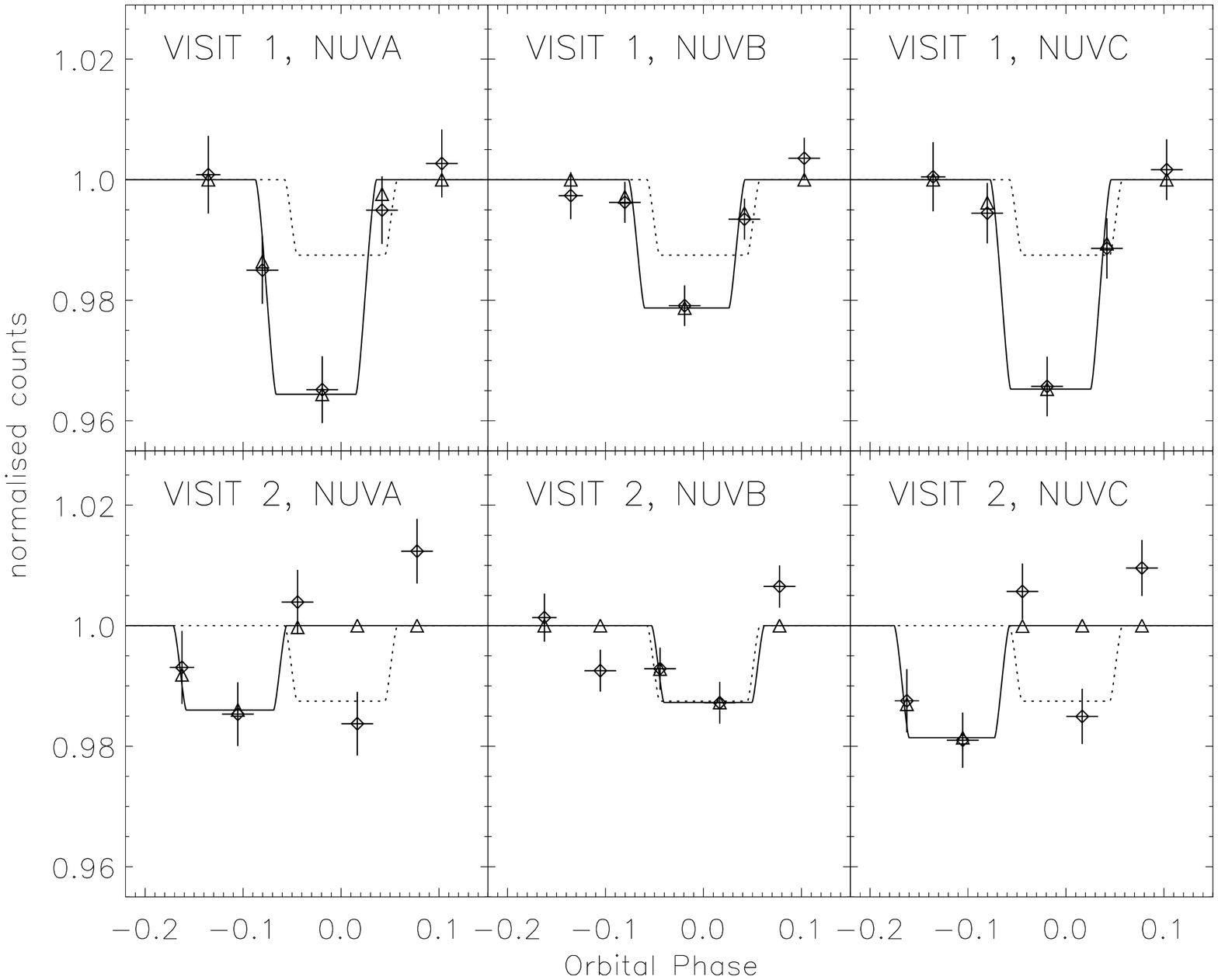}
\caption{The best-fitting simple transit models (solid lines) for each of our
NUV light curves (diamonds). Triangles show the integral of the
best-fit model over the exposure times used.
In Visit 1 we obtain good fits which reveal an
NUV transit deeper than the optical transit, and an early ingress.
Visit 2 is less clear-cut: the 3rd data point in all three bands
is high compared to the models fitting Visit 1, particularly for NUVA
and NUVC. For these two bands the best fit model places this
anomalously high point after egress so the `best-fit' transit
does not overlap the optical transit (dotted line).
}
\label{fig:1visfits}
\end{center}
\end{figure}

 %\label{table_chi}
In Paper 1 we observed an NUV transit which was significantly deeper than
the optical transit, and which began with an early ingress compared to
the optical ephemeris. These conclusions from our Visit 1 data alone
are strengthened by the re-reduction described in \S\ref{hstobs}.
We quantify this by fitting the simplest possible transit model,
assuming a circular occulting disc transiting a uniform brightness (no
limb-darkening) star. We fix all the parameters of this model to the values demanded
by WASP-12\,b's optical transit except for two free parameters: $R/R_{*} \ge R_{\rm P}/R_{*}$, and
$\Delta \phi$, describing the radius and phase offset of the NUV occulting disc compared to
the transit of the optically opaque planet with $R=R_{\rm P}$ and $\Delta \phi= 0$ across the star of radius $R_*$.
The normalisation of the data to the model is also a free parameter.
We report our model fitting in Table 4, using the reduced-$\chi^2$ statistic, $\chi^2_{\nu}$,
which is the appropriate statistic for assessing the goodness of fit.
For completeness
we also report the number of degrees of freedom, i.e. the number of parameters including normalisation,
minus the number of fitted data points.
The results for Visit 1 shown in Fig.~\ref{fig:chi_nuv} and at
the top of Fig.~\ref{fig:1visfits} and Table~\ref{table_chi}
are consistent with Paper 1:
for NUVA, NUVB, and NUVC the best-fit transits are early by 43~min, 29~min, and 26~min
respectively. As Table~\ref{table_chi} shows, the best-fit radii for the occulting
disc are significantly enhanced compared to that of the optically opaque planet \citep[$R_{\rm P}/R_{*}=0.1119$,][]{Chan11}
with an enhancement in radius of $70\%$, i.e. almost a factor of three in area,
for the NUVA and NUVC bands.
The NUVA and NUVC bands provide extremely sensitive probes for the presence of low density
gas at temperatures of several thousand degrees because these bands contain $\sim 10^3$
overlapping spectral lines. We discuss the spectral information in detail in
\S\ref{sec:spec-interp} below.
We do not imagine that the low density gas we have detected surrounding and preceding
WASP-12\,b actually presents a circular cross-section to us during its transit;
this was the simplest model we could imagine, and we have few data points to fit.
As the upper panels in Fig.~\ref{fig:1visfits} show, with two free parameters (three including normalisation)
the model is already able to produce an improbably good fit to the data, and this suggests that our
error bars (derived from photon-counting statistics) are not
under-estimated.

%--------------------------------------------------------------------
%--------------------------------------------------------------------

%--------------------------------------------------------------------
%--------------------------------------------------------------------

Visit 2 was timed to provide coverage of the orbital phases missing due to
Earth-occultation in our Visit 1 data. The two visits provide well-sampled
coverage between $ -0.17 < \phi_{\rm orb} < 0.12 $.
%{\bf fill in last sig fig after ephem update and Luca's obs tab}.
Unfortunately the Visit 2 data defy straightforward interpretation.
The lower panels of Fig.~\ref{fig:1visfits} show the Visit 2 data along
with the best-fit simple transit models to the three NUV bands. As the corresponding
entries in Table~\ref{table_chi} show, none of these fits are
formally acceptable, with $\chi^2_{\nu}$ significantly above 1 in all cases.
The NUVB band produces the best fit transit for Visit 2, but this fit essentially
reproduces the transit of the optically opaque planet (dotted line in
Fig.~\ref{fig:1visfits}), with points before and after the transit deviating significantly
from the model. The best-fit NUVB model is slightly late compared to the optical transit because the Visit 2 point which samples optical ingress is higher than the optical transit would predict.
The Visit 2 NUVA and NUVC light curves strongly resemble each other. Throughout, the
point-to-point deviations in the two light curves appear similar.
In both cases the Visit 2 data is strongly absorbed before optical first contact,
with the point sampling optical ingress lying above the out of transit level, while
the Visit 1 data appears more strongly absorbed between optical second and third contacts.
The best-fitting model for the Visit 2 NUVA and NUVC light curves is a transit which occurs so early that it
fails to overlap the optical transit. This is clearly unphysical: the planet
is surely opaque to NUV light. These fits result from the anomalously high third data
point (arising from the third HST orbit) in the Visit 2 NUVA and NUVC light curves: in both cases the exposure spanning the
optical ingress is the 2nd-highest in the light curve. Our simple model requires a
strict relationship between the depth and duration of the fitted transit. Since the `orbit 3 point'
 is high, the model
places it above the out-of-transit level and as the
pre-optical-ingress
NUVA and NUVC data points are both lower than the fitted out of transit level, the best-fit
transit model places the entire transit early.

The aberrantly high Visit 2 orbit 3 data point in the NUVA and NUVC light curves thus defies the simple explanation
we developed from Visit 1.  We can think of three plausible explanations
for this: (i) these are noisy data and this simply happens to be a point with a positive deviation; (ii) there
was a short-lived stellar flare during Visit 2 orbit 3; (iii) the gas which absorbs before
first contact is clumpy, and there was a clear window to the star during  Visit 2 orbit 3.
\citet{vidotto11}'s bow shock model can produce such a `window' configuration.
The same pattern independently appearing in both the NUVA and NUVC light curves makes the first explanation far less likely than it would otherwise be, and our fits to the Visit 1 light curve suggest our error estimations are ample.
We will discuss the stellar flare hypothesis in \S\S\ref{sec:abs} and ~\ref{sec:flare}, and the window hypothesis in \S\ref{sec:flare}.

%--------------------------------------------------------------------
\begin{figure}
\begin{center}
\includegraphics[width=17cm]{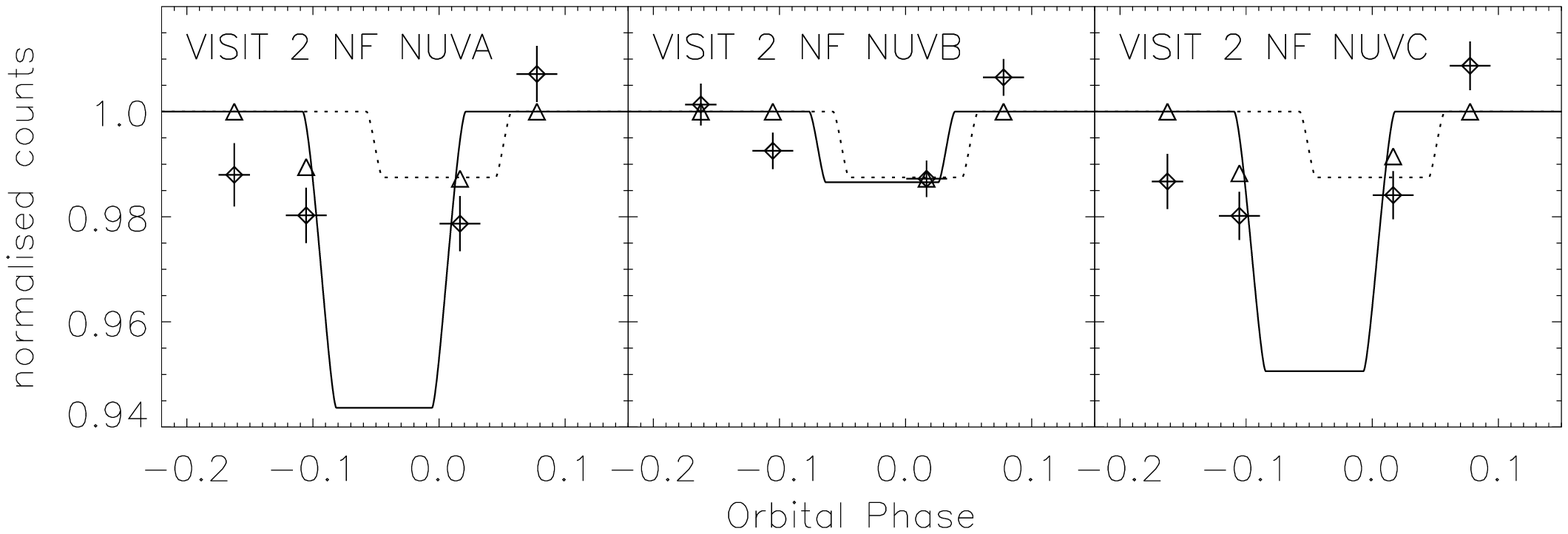}
\caption{The fits which result for Visit 2 if we mask out the aberrant third
data point. The NUVA and NUVC fits have absurd depths because the
observed transit has a long duration and we fit a circular
occulter. A more plausible model would have absorbing gas
extended in the orbital plane. Symbols as in Fig.~\ref{fig:1visfits}.
}
\label{fig:vis2_NF_fits}
\end{center}
\end{figure}
%--------------------------------------------------------------------

The Visit 2 data with the aberrant 3rd-orbit
data masked out produces the fits shown in Fig~\ref{fig:vis2_NF_fits} and Table~\ref{table_chi}
(entries for visit `2NF').
The NUVB fit now has a slightly deeper transit which occurs earlier because the fit has placed the orbit 4 point to span the beginning of egress.  This fit, and the adjustments to it we might
make if we relax our arbitrary assumption that the extended gas presents a circular
cross-section, is consistent with the interpretation we made in Paper 1. The NUVA and NUVC
`best' fits are comical. Because the orbit 1, 2, and 4 points in both cases lie so far
below the orbit 5 point, the relationship between transit depth and duration favors
fits in which the transit depth is much deeper than any data point. A more sensible
interpretation would be that the absorbing gas is extended along the orbital plane and presents
a significantly non-circular cross-section, so the duration is longer than that of the models while the
depth is shallower.

The fits we have performed separately on the two individual visits suggest that the configuration
of the absorbing gas changed between the two visits. Indeed Vidotto, Jardine \& Helling (2011)
% MNRAS, 414, 1573
interpret this variability.
Nonetheless for completeness we now proceed to examine the data from the two visits together.

\begin{figure}
\begin{center}
\includegraphics[width=100mm]{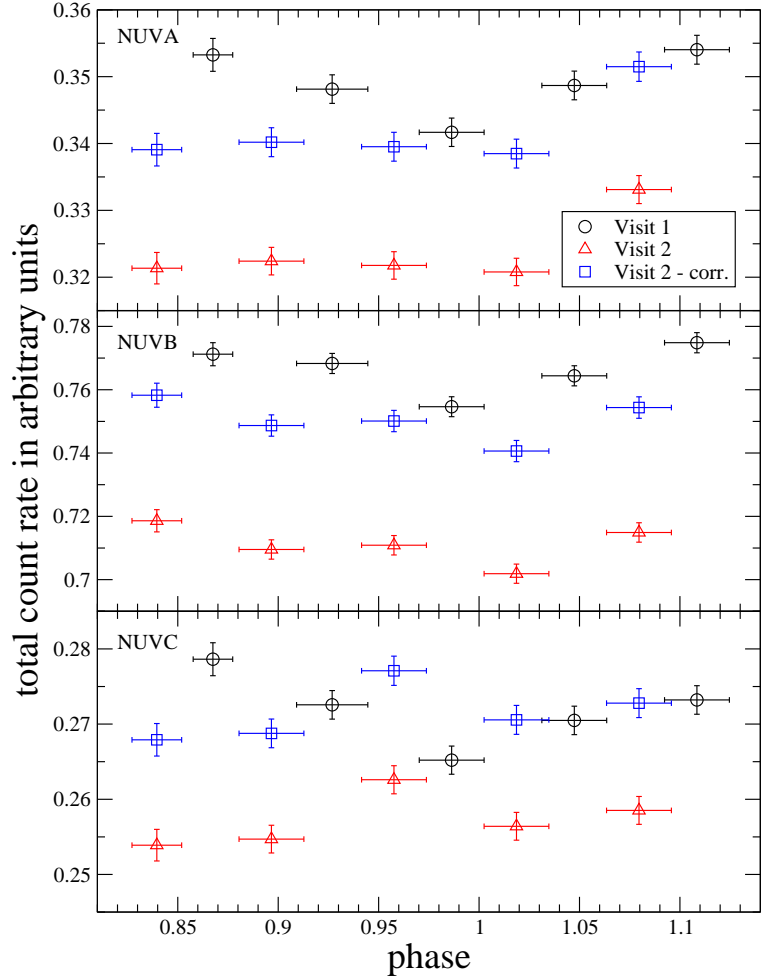}
\caption{Count rates from the NUVA, NUVB, and NUVC overlapping spectral regions in
the two HST visits.  Circles show Visit 1, triangles
show Visit 2, and squares show Visit 2 after multiplying by 1.052 to correct for
the decline in instrument sensitivity between the two visits.
}
\label{nonorm-compv1v2}
\end{center}
\end{figure}

Fig.~\ref{nonorm-compv1v2} interleaves the two visits using only the photons detected
within the overlapping wavelength regions
of the RV-corrected spectra.
It is immediately apparent that the Visit 1 data (black circles) lies
above the Visit 2 data (red triangles). We attribute this
to the known declining throughput of the COS/G285M
configuration \citep{Osten_ISR11-02}.
We corrected for this by boosting
the Visit 2 count-rates by 5.52\% using an
estimate from  \citet{Osten},
but  the
rate of sensitivity decline is not precisely known for our instrument configurations.
The `corrected' Visit 2 data are shown as blue squares.
%\footnote{
%http://www.stsci.edu/hst/cos/documents/newsletters/full\_stories/2010\_05/fuv\_sens\_loss.html}
%--------------------------------------------------------------------
%--------------------------------------------------------------------

The background level differed between the two visits, with roughly twice as many background counts
during Visit 2. The temporal behaviour of the background also differed, with a gradual increase in
background during Visit 1 contrasting with a decrease in background during Visit 2. We
corrected for background, so this should not affect our light curves. The corrected Visit 2 data
appear to match well at phases following the transit for NUVA and NUVC, but the corrected
NUVB light curve
from Visit 2 appears systematically lower than the Visit 1 light curve.
The Visit 2 orbit 3 point remains aberrant in NUVC, but restricting the wavelength coverage
to only the overlapping region has dramatically lowered the NUVA orbit 3 point. We will return
to discuss the reasons for this in \S\ref{sec:flare}.

\begin{figure}
\begin{center}
\includegraphics[width=100mm]{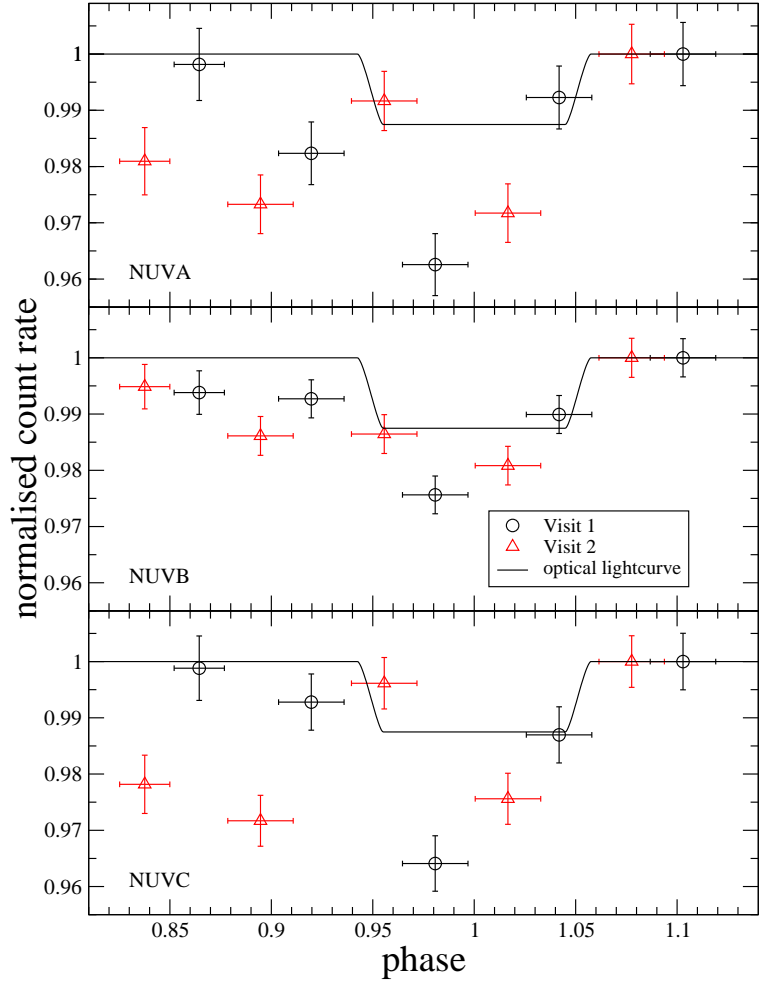}
\caption{The light curves from all of our HST data, with each HST visit
normalised to unity in the last HST orbit.
Note the wavelength regions
contributing to the two visits differ, as indicated in Figure~\ref{flux-v1ev2}.
}
\label{nonorm-v1v2}
\end{center}
\end{figure}

Our NUV light curves, particularly the NUVA and NUVC bands, suggest
that in Visit 2 perhaps the  absorbing gas obscured the
stellar disc at phases as early as $\phi_{\rm orb} = 0.83$, in
which case Visit 2 contains no genuine pre-transit data.
Fig.~\ref{nonorm-v1v2} shows
the light curves produced using all the data (i.e. not just the overlapping wavelength regions),
arbitrarily
setting the last point in each visit to (an out-of-transit level of) 1.0.
The NUVB light curve now looks perfectly consistent with the interpretation
we made in Paper 1: the transit is deeper than the optical
transit and appears to be preceded by absorption occurring before
optical ingress. The NUVA and NUVC light curves also look reasonably consistent with
this interpretation, with much more significant pre-ingress absorption
occurring in Visit 2.

%--------------------------------------------------------------------
%--------------------------------------------------------------------

\begin{figure}
\begin{center}
\includegraphics[width=17cm]{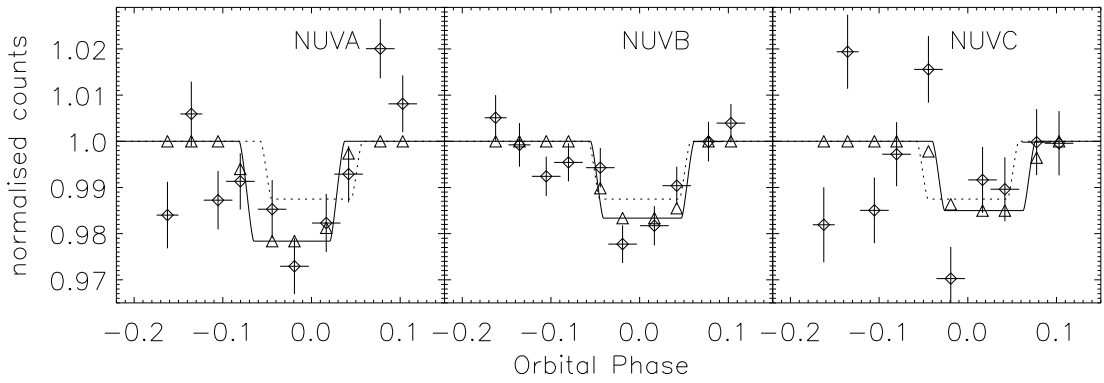}
\caption{The best-fitting simple transit models for our interleaved
NUV data. None of the fits are formally acceptable, but
the NUVA and NUVB results are broadly consistent with a deeper
transit than the optical and absorption from diffuse gas which
precedes the optical transit (dotted line). The best-fit NUVC transit
occurs late, but the first half of the interleaved lightcurve
has extremely high scatter so this fit should clearly not be taken
at face value. Symbols as in Fig.~\ref{fig:1visfits}.
}
\label{fig:olap_fits}
\end{center}
\end{figure}

We performed simple transit fits (as described above) to the combined data
from the overlapping wavelength regions in the two visits, allowing
the normalisation of each visit to vary independently. These fits thus have
four free parameters in total: $R$, $\Delta \phi$, and two
independent normalisation values. Fig~\ref{fig:olap_fits} shows the best fitting
models, along with the data after applying the best fit normalisations.
As Table~\ref{table_chi} quantifies (entries for bands AO, BO,
CO; Visits `1,2'), none of the
three fits in Fig.~\ref{fig:olap_fits} are formally acceptable. Nonetheless, the NUVA and NUVB
fits are broadly consistent with our interpretation in Paper 1. The best-fit
transit is significantly deeper than the optical transit and the light curves show signs
of absorption from diffuse gas before optical ingress.
The NUVC band produces formally the worst fit, and the first half of the light curve
shows enormous scatter between the sets of points from the two visits. The best
fit model transit is late compared to the optical transit: a fit which moved the ingress
to after the anomalously high Visit 2 orbit 3 point has been preferred.
%--------------------------------------------------------------------
%--------------------------------------------------------------------

%--------------------------------------------------------------------
\begin{figure}
\begin{center}
\includegraphics[width=17cm]{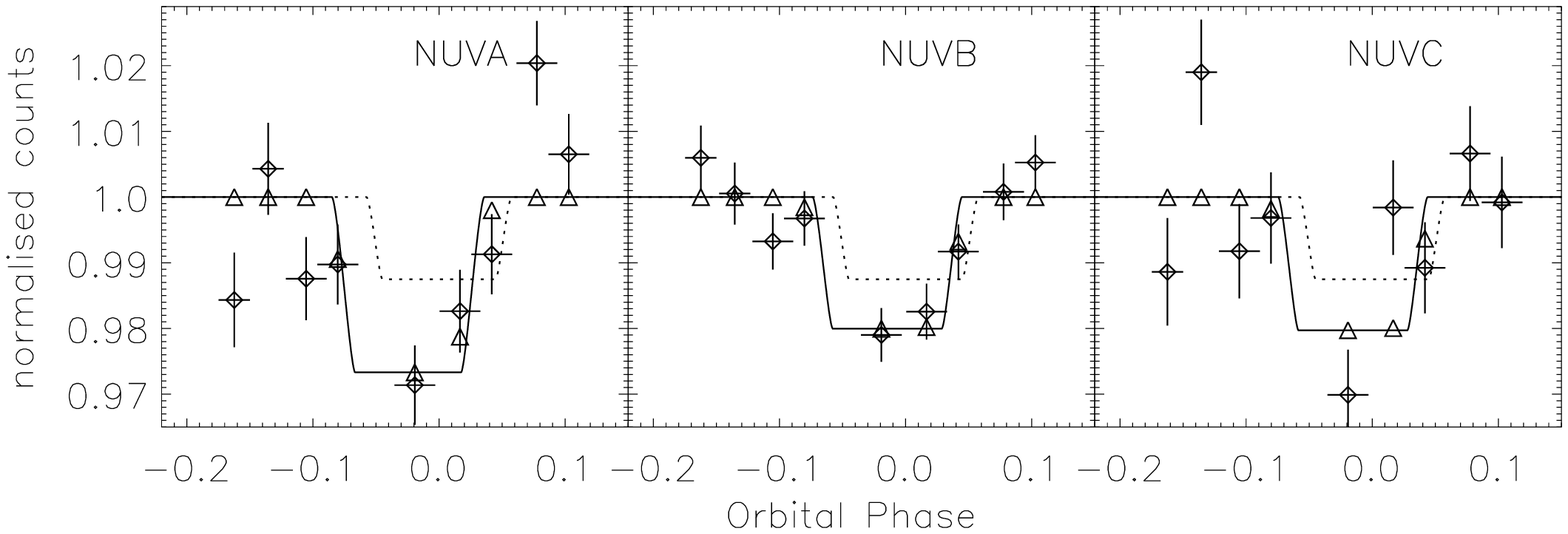}
\caption{The best-fitting simple transit models for our interleaved
NUV data with the Visit 2 Orbit 3 point masked out. None of the fits are formally acceptable, but
the results are broadly consistent with a deeper
transit than the optical and absorption from  gas
preceding optical transit (dotted line). Symbols as in Fig.~\ref{fig:1visfits}.
}
\label{fig:olap_NF_fits}
\end{center}
\end{figure}
%--------------------------------------------------------------------

Figure~\ref{fig:olap_NF_fits} shows the best fits to the interleaved data
without the Visit 2 orbit 3 point. These fits (entries for Visits `1,2NF')
are all broadly consistent
with our Paper 1 conclusions. For the NUVB and NUVC bands, the best $\chi^2_\nu$
is better than the fits shown in Fig.~\ref{fig:olap_fits}.

In both NUVA and NUVC, as Figure~\ref{flux-v1ev2} shows, the wavelength ranges covered only by the Visit 1
data are strongly absorbed by the stellar photosphere, while the wavelength ranges covered only by
the Visit 2 data are significantly less strongly absorbed.
If the diffuse absorbing  gas has broadly the same ionic composition (i.e. abundance mix and temperature)
as the stellar photosphere, this would predict a deeper transit in Visit 1 NUVA and NUVC than in Visit 2 NUVA and NUVC.
This is exactly what Fig.~\ref{fig:1visfits} and Table~\ref{table_chi} show.

Figure~\ref{flux-v1ev2} shows that the
NUVB spectral region, for both
visits, is the closest to the stellar continuum: there is relatively
little photospheric absorption, and we might similarly expect relatively little absorption
from the extended cloud of diffuse gas. Of course, without knowing the composition and
physical conditions of this gas, it is impossible to assert this is definitely true,
but it is the simplest assumption one could make {\it a priori}. Our NUVB
data suggests a deeper transit than that of the optical light curve, but of the three
spectral regions it is the closest to the optical curve.
Note the NUVB data have the smallest uncertainties
because, being relatively unabsorbed, we detect most photons from the star in NUVB.

Each (version of the) NUVB light curve (Figs.~\ref{fig:1visfits}, ~\ref{fig:vis2_NF_fits}, ~\ref{nonorm-compv1v2},
~\ref{nonorm-v1v2}, ~\ref{fig:olap_fits} and ~\ref{fig:olap_NF_fits})
shows indications of an early ingress in NUVB. Our NUVA and NUVC light curves are also broadly consistent with this, except for the anomalous count rates
observed in Visit 2 orbit 3. This aberrant point appears in both NUVA and NUVC when we considered the entire
wavelength range (Figs.~\ref{fig:1visfits} and ~\ref{nonorm-v1v2}) but only in NUVC
when we extract light curves from the overlapping wavelength ranges (Figs.~\ref{nonorm-compv1v2} and ~\ref{fig:olap_fits}).
The Visit 2 orbit 3 NUVA point shifts downwards
by $ 4\sigma$ when we restrict the wavelength range. We think this is significant and will return to discuss the implications
of this piece of evidence in \S\ref{sec:flare}.

In summary, the NUV light curves are broadly consistent with our
principal conclusions
in Paper 1. The transit is deeper than in the optical, and is preceded by an early ingress, though
the point for the
third
orbit in Visit 2 deviates from this description when some wavelength regions are considered.
There is clearly NUV absorption
occurring before optical first contact.

\subsection{The wavelength resolved transit}
\label{sec:spec-interp}
The enhanced transit depths seen in Figs.~\ref{fig:chi_nuv},~\ref{fig:1visfits},~\ref{fig:vis2_NF_fits},
\ref{nonorm-v1v2},~\ref{fig:olap_fits} and \ref{fig:olap_NF_fits}
result from summing over the $\sim 40 \AA$ covered in each of our NUVA, NUVB and NUVC spectra (or the $\sim 30\AA$ of
overlapping wavelength coverage for the last two figures listed).
The enhancement of the NUV transit depth compared to the optical transit depth
is due to absorption in the diffuse gas surrounding the planet. This
absorption is caused by bound-bound transitions in
atomic and ionic species. {\it A priori} we do not know which species are
present in this gas, but we do know that there are thousands of
overlapping spectral lines within the wavelength ranges we have observed.
Since we have time-series spectral data, we can examine the transit at
any spectral resolution greater than or equal to that of the data.
In Paper 1 we presented evidence for detections of enhanced transit
depths in \ion{Mg}{2} and other lines of neutral and ionised metals.
With the addition of Visit 2 we have extended our wavelength coverage,
our temporal coverage, and increased the SNR of the overlapping
wavelength region, so we examined our Visit 1 and Visit 2 data afresh
for evidence of enhanced transit depths at particular
wavelengths.

{\tiny
\onecolumn
\begin{longtable}{ll|ll||ll|ll||ll|ll}
%\begin{table}
\caption{{\small {\bf online material} Wavelength of the Visit 1 and Visit 2 spectral points deviating more
than 3$\sigma$, adopting two different $\sigma$s: $\sigma_{d_{\lambda}|_{exp}}$ and
$\sigma_{d_{\lambda}|_{prop}}$. The deviating points highlighted in bold face are the Visit 2
points which are one resolution element away from a Visit 1 point. All wavelengths are given in \AA.}
}
\label{tab:lam_id}
%\begin{tabular}{ll|ll||ll|ll||ll|ll}
\protect\label{all-lines}\\
\tableline\tableline \\
Wavelength                     & Wavelength                     & Wavelength                      & Wavelength                      & Wavelength                     & Wavelength                     & Wavelength                      & Wavelength                      & Wavelength                     & Wavelength                     & Wavelength                      & Wavelength                      \\
$\lambda - 2000$\AA            & $\lambda - 2000$\AA            & $\lambda - 2000$\AA             & $\lambda - 2000$\AA             & $\lambda - 2000$\AA            & $\lambda - 2000$\AA            & $\lambda - 2000$\AA             & $\lambda - 2000$\AA             & $\lambda - 2000$\AA            & $\lambda - 2000$\AA            & $\lambda - 2000$\AA             & $\lambda - 2000$\AA             \\
Visit 1                        & Visit 2                        & Visit 1                         & Visit 2                         & Visit 1                        & Visit 2                        & Visit 1                         & Visit 2                         & Visit 1                        & Visit 2                        & Visit 1                         & Visit 2                        \\
3$\times\sigma_{d_{i}|_{exp}}$ & 3$\times\sigma_{d_{i}|_{exp}}$ & 3$\times\sigma_{d_{i}|_{prop}}$ & 3$\times\sigma_{d_{i}|_{prop}}$ & 3$\times\sigma_{d_{i}|_{exp}}$ & 3$\times\sigma_{d_{i}|_{exp}}$ & 3$\times\sigma_{d_{i}|_{prop}}$ & 3$\times\sigma_{d_{i}|_{prop}}$ & 3$\times\sigma_{d_{i}|_{exp}}$ & 3$\times\sigma_{d_{i}|_{exp}}$ & 3$\times\sigma_{d_{i}|_{prop}}$ & 3$\times\sigma_{d_{i}|_{prop}}$ \\
\multicolumn{4}{c||}{NUVA} & \multicolumn{4}{|c||}{NUVB} & \multicolumn{4}{c}{NUVC} \\
\tableline
\endfirsthead
\caption{continued.}\\
\tableline\tableline \\
Wavelength                     & Wavelength                     & Wavelength                      & Wavelength                      & Wavelength                     & Wavelength                     & Wavelength                      & Wavelength                      & Wavelength                     & Wavelength                     & Wavelength                      & Wavelength                      \\
$\lambda - 2000$\AA            & $\lambda - 2000$\AA            & $\lambda - 2000$\AA             & $\lambda - 2000$\AA             & $\lambda - 2000$\AA            & $\lambda - 2000$\AA            & $\lambda - 2000$\AA             & $\lambda - 2000$\AA             & $\lambda - 2000$\AA            & $\lambda - 2000$\AA            & $\lambda - 2000$\AA             & $\lambda - 2000$\AA             \\
Visit 1                        & Visit 2                        & Visit 1                         & Visit 2                         & Visit 1                        & Visit 2                        & Visit 1                         & Visit 2                         & Visit 1                        & Visit 2                        & Visit 1                         & Visit 2                        \\
3$\times\sigma_{d_{i}|_{exp}}$ & 3$\times\sigma_{d_{i}|_{exp}}$ & 3$\times\sigma_{d_{i}|_{prop}}$ & 3$\times\sigma_{d_{i}|_{prop}}$ & 3$\times\sigma_{d_{i}|_{exp}}$ & 3$\times\sigma_{d_{i}|_{exp}}$ & 3$\times\sigma_{d_{i}|_{prop}}$ & 3$\times\sigma_{d_{i}|_{prop}}$ & 3$\times\sigma_{d_{i}|_{exp}}$ & 3$\times\sigma_{d_{i}|_{exp}}$ & 3$\times\sigma_{d_{i}|_{prop}}$ & 3$\times\sigma_{d_{i}|_{prop}}$ \\
\multicolumn{4}{c||}{NUVA} & \multicolumn{4}{|c||}{NUVB} & \multicolumn{4}{c}{NUVC} \\
\tableline
\endhead
\tableline
\endfoot
\tableline
\tableline
\endlastfoot
538.484       &               & 537.535       &               & 661.605       &               & 657.054       &               &               & 792.434       & 776.577       &               \\
539.002       & 	      & 537.880       & 	      & 662.022       & 	      & 664.316       & 	      & 793.963       & 	      & 776.818       & 	      \\
539.045       & 	      & 538.139       & 	      & 663.649       & 	      & 664.816       & 	      & 794.442       & 	      & 778.785       & 	      \\
543.442       & 	      & 538.312       & 	      & 666.649       & 	      & 666.691       & 	      & 794.842       & 	      & 779.026       & 	      \\
546.758       & 	      & 538.916       & 	      &               & 667.763       & 666.733       & 	      & 794.922       & 	      & 784.399       & 	      \\
557.028       & 	      & 538.959       & 	      &               & 667.888       &               & 667.763       & 795.401       & 	      & 785.040       & 	      \\
562.344       & 	      & 539.520       & 	      & 669.522       & 	      &               & 669.432       & 795.561       & 	      &               & 787.704       \\
562.473       & 	      & 539.606       & 	      & 670.313       & 	      &               & 669.474       & 795.641       & 	      &               & 787.945       \\
              & 563.215       & 540.166       & 	      &               & 671.851       &               & 672.310       & 797.957       & 	      &               & 788.065       \\
563.501       & 	      & 540.641       & 	      &               & 672.018       &               & 678.558       & 801.428       & 	      & 788.524       &               \\
              & {\bf 575.914} & 540.856       & 	      &               & {\bf 678.807} &               & 680.804       & 802.185       & 	      &               & {\bf 790.952} \\
{\bf 575.980} &               & 540.900       & 	      & {\bf 678.956} &               & 684.179       & 	      & 802.425       & 	      & {\bf 791.004} & 	      \\
              & {\bf 575.999} & 541.072       & 	      & 682.895       & 	      &               & 686.995       & 802.504       & 	      & {\bf 791.084} & 	      \\
{\bf 576.151} & 	      & 541.158       & 	      & 682.978       & 	      & 687.491       & 	      & 803.023       & 	      & 791.164       & 	      \\
              & {\bf 576.342} & 541.201       & 	      & 684.759       & 	      &               & 687.908       & 803.142       & 	      &               & {\bf 791.513} \\
              & 576.385       & 541.891       & 	      &               & 689.360       & 690.056       & 	      & 	      & 	      & {\bf 791.604} & 	      \\
              & 576.470       & 542.107       & 	      &               & 690.936       & 691.089       &               & 	      & 	      &               & 792.274       \\
              & 577.883       & 542.193       & 	      &               & 690.978       &               & {\bf 692.470} & 	      & 	      &               & 792.314       \\
580.925       & 	      & 542.537       & 	      & 695.261       & 	      & {\bf 692.618} & 	      & 	      & 	      &               & 792.674       \\
              & 582.286       & 542.581       & 	      &               & 698.681       &               & {\bf 692.802} & 	      & 	      & {\bf 793.123} & 	      \\
              & 585.873       & 543.873       & 	      &               & 698.805       & 694.601       & 	      & 	      & 	      &               & {\bf 793.235} \\
              & 586.043       & 543.916       & 	      &               & 698.847       & {\bf 695.261} & 	      & 	      & 	      & {\bf 793.243} & 	      \\
	      & 	      & 544.347       & 	      &               & 706.904       &               & {\bf 695.370} & 	      & 	      & {\bf 793.403} & 	      \\
	      & 	      & 544.390       & 	      &               & 710.368       &               & 695.784       & 	      & 	      & 793.443       & 	      \\
	      & 	      & 544.433       & 	      & 	      & 	      &               & 697.233       & 	      & 	      & {\bf 793.483} & 	      \\
	      & 	      & 546.155       & 	      & 	      & 	      &               & 700.873       & 	      & 	      &               & {\bf 793.635} \\
	      & 	      & 546.198       & 	      & 	      & 	      &               & 703.146       & 	      & 	      &               & {\bf 793.715} \\
	      & 	      & 546.241       & 	      & 	      & 	      &               & 706.408       & 	      & 	      & {\bf 793.723} & 	      \\
	      & 	      & 546.715       & 	      & 	      & 	      &               & 710.080       & 	      & 	      &               & {\bf 793.755} \\
	      & 	      & 547.490       & 	      & 	      & 	      &               & 710.203       & 	      & 	      & {\bf 793.803} & 	      \\
	      & 	      & 547.705       & 	      & 	      & 	      &               & 710.327       & 	      & 	      & {\bf 793.843} & 	      \\
	      & 	      & 548.952       & 	      & 	      & 	      &               & 710.368       & 	      & 	      & 794.003       & 	      \\
	      & 	      & 549.425       & 	      & 	      & 	      & 	      & 	      & 	      & 	      & {\bf 794.083} & 	      \\
	      & 	      &               & {\bf 549.902} & 	      & 	      & 	      & 	      & 	      & 	      &               & {\bf 794.116} \\
	      & 	      & {\bf 549.984} &               & 	      & 	      & 	      & 	      & 	      & 	      & {\bf 794.402} & 	      \\
	      & 	      & {\bf 550.027} & 	      & 	      & 	      & 	      & 	      & 	      & 	      &               & {\bf 794.436} \\
	      & 	      &               & 550.248       & 	      & 	      & 	      & 	      & 	      & 	      & {\bf 794.522} & 	      \\
	      & 	      &               & 550.291       & 	      & 	      & 	      & 	      & 	      & 	      &               & 794.956       \\
	      & 	      &               & 550.550       & 	      & 	      & 	      & 	      & 	      & 	      &               & 794.996       \\
	      & 	      &               & 550.636       & 	      & 	      & 	      & 	      & 	      & 	      &               & 795.196       \\
	      & 	      &               & 550.680       & 	      & 	      & 	      & 	      & 	      & 	      &               & 795.276       \\
	      & 	      & 550.887       & 	      & 	      & 	      & 	      & 	      & 	      & 	      &               & 795.316       \\
	      & 	      &               & 551.629       & 	      & 	      & 	      & 	      & 	      & 	      &               & 795.356       \\
	      & 	      &               & 551.758       & 	      & 	      & 	      & 	      & 	      & 	      &               & 795.397       \\
	      & 	      &               & 551.802       & 	      & 	      & 	      & 	      & 	      & 	      &               & {\bf 795.477} \\
	      & 	      &               & 552.708       & 	      & 	      & 	      & 	      & 	      & 	      & {\bf 795.481} & 	      \\
	      & 	      &               & 553.010       & 	      & 	      & 	      & 	      & 	      & 	      &               & {\bf 795.517} \\
	      & 	      &               & 553.139       & 	      & 	      & 	      & 	      & 	      & 	      & {\bf 795.521} & 	      \\
	      & 	      & 553.465       & 	      & 	      & 	      & 	      & 	      & 	      & 	      &               & {\bf 795.597} \\
	      & 	      &               & 554.131       & 	      & 	      & 	      & 	      & 	      & 	      & {\bf 795.601} & 	      \\
	      & 	      &               & 554.691       & 	      & 	      & 	      & 	      & 	      & 	      &               & {\bf 795.637} \\
	      & 	      &               & 555.424       & 	      & 	      & 	      & 	      & 	      & 	      & {\bf 795.721} & 	      \\
	      & 	      &               & 555.510       & 	      & 	      & 	      & 	      & 	      & 	      & 795.920       & 	      \\
	      & 	      &               & 556.975       & 	      & 	      & 	      & 	      & 	      & 	      & 796.080       & 	      \\
	      & 	      &               & 557.277       & 	      & 	      & 	      & 	      & 	      & 	      & 796.120       & 	      \\
	      & 	      &               & 557.578       & 	      & 	      & 	      & 	      & 	      & 	      & 796.160       & 	      \\
	      & 	      &               & 558.310       & 	      & 	      & 	      & 	      & 	      & 	      & 796.200       & 	      \\
	      & 	      &               & 558.396       & 	      & 	      & 	      & 	      & 	      & 	      & {\bf 796.280} & 	      \\
	      & 	      &               & {\bf 558.870} & 	      & 	      & 	      & 	      & 	      & 	      &               & {\bf 796.357} \\
	      & 	      & {\bf 558.916} & 	      & 	      & 	      & 	      & 	      & 	      & 	      & {\bf 796.479} & 	      \\
	      & 	      & 559.173       & 	      & 	      & 	      & 	      & 	      & 	      & 	      &               & {\bf 796.557} \\
	      & 	      &               & 559.387       & 	      & 	      & 	      & 	      & 	      & 	      &               & {\bf 796.597} \\
	      & 	      &               & 559.430       & 	      & 	      & 	      & 	      & 	      & 	      &               & {\bf 796.637} \\
	      & 	      &               & {\bf 559.774} & 	      & 	      & 	      & 	      & 	      & 	      & {\bf 796.639} & 	      \\
	      & 	      & {\bf 559.816} & 	      & 	      & 	      & 	      & 	      & 	      & 	      & {\bf 796.679} & 	      \\
	      & 	      & {\bf 559.859} & 	      & 	      & 	      & 	      & 	      & 	      & 	      &               & {\bf 796.717} \\
	      & 	      &               & 560.075       & 	      & 	      & 	      & 	      & 	      & 	      &               & {\bf 796.797} \\
	      & 	      &               & 560.549       & 	      & 	      & 	      & 	      & 	      & 	      &               & {\bf 796.917} \\
	      & 	      &               & 561.237       & 	      & 	      & 	      & 	      & 	      & 	      & {\bf 796.919} & 	      \\
	      & 	      &               & 561.280       & 	      & 	      & 	      & 	      & 	      & 	      &               & {\bf 796.957} \\
	      & 	      &               & 561.796       & 	      & 	      & 	      & 	      & 	      & 	      &               & {\bf 797.197} \\
	      & 	      &               & 561.925       & 	      & 	      & 	      & 	      & 	      & 	      & {\bf 797.318} & 	      \\
	      & 	      &               & 562.312       & 	      & 	      & 	      & 	      & 	      & 	      & {\bf 797.358} & 	      \\
	      & 	      & {\bf 562.558} & 	      & 	      & 	      & 	      & 	      & 	      & 	      & 797.438       & 	      \\
	      & 	      &               & {\bf 562.570} & 	      & 	      & 	      & 	      & 	      & 	      & 797.478       & 	      \\
	      & 	      & {\bf 562.601} & 	      & 	      & 	      & 	      & 	      & 	      & 	      & {\bf 797.518} & 	      \\
	      & 	      &               & 562.871       & 	      & 	      & 	      & 	      & 	      & 	      &               & {\bf 797.717} \\
	      & 	      &               & 562.914       & 	      & 	      & 	      & 	      & 	      & 	      & {\bf 797.797} & 	      \\
	      & 	      & {\bf 563.372} & 	      & 	      & 	      & 	      & 	      & 	      & 	      & {\bf 797.837} & 	      \\
	      & 	      &               & {\bf 563.387} & 	      & 	      & 	      & 	      & 	      & 	      & {\bf 797.917} & 	      \\
	      & 	      & {\bf 563.415} & 	      & 	      & 	      & 	      & 	      & 	      & 	      & {\bf 797.997} & 	      \\
	      & 	      &               & {\bf 563.430} & 	      & 	      & 	      & 	      & 	      & 	      &               & {\bf 798.157} \\
	      & 	      &               & 563.516       & 	      & 	      & 	      & 	      & 	      & 	      &               & 798.517       \\
	      & 	      &               & 563.645       & 	      & 	      & 	      & 	      & 	      & 	      &               & {\bf 798.756} \\
	      & 	      &               & 563.731       & 	      & 	      & 	      & 	      & 	      & 	      & {\bf 798.875} & 	      \\
	      & 	      &               & 564.161       & 	      & 	      & 	      & 	      & 	      & 	      & 799.114       & 	      \\
	      & 	      &               & 564.548       & 	      & 	      & 	      & 	      & 	      & 	      & 799.633       & 	      \\
	      & 	      &               & 564.762       & 	      & 	      & 	      & 	      & 	      & 	      &               & 799.956       \\
	      & 	      &               & 565.063       & 	      & 	      & 	      & 	      & 	      & 	      &               & 800.355       \\
	      & 	      &               & 566.824       & 	      & 	      & 	      & 	      & 	      & 	      &               & 800.395       \\
	      & 	      &               & 566.996       & 	      & 	      & 	      & 	      & 	      & 	      &               & 800.475       \\
	      & 	      & 567.823       & 	      & 	      & 	      & 	      & 	      & 	      & 	      &               & 800.515       \\
	      & 	      &               & 568.026       & 	      & 	      & 	      & 	      & 	      & 	      &               & {\bf 800.675} \\
	      & 	      & 568.635       & 	      & 	      & 	      & 	      & 	      & 	      & 	      &               & {\bf 800.715} \\
	      & 	      &               & 569.314       & 	      & 	      & 	      & 	      & 	      & 	      & {\bf 800.790} & 	      \\
	      & 	      &               & 569.700       & 	      & 	      & 	      & 	      & 	      & 	      &               & 800.994       \\
	      & 	      &               & 569.828       & 	      & 	      & 	      & 	      & 	      & 	      &               & {\bf 801.434} \\
	      & 	      &               & 569.957       & 	      & 	      & 	      & 	      & 	      & 	      & {\bf 801.548} & 	      \\
	      & 	      & 570.857       & 	      & 	      & 	      & 	      & 	      & 	      & 	      &               & {\bf 801.634} \\
	      & 	      & 570.900       & 	      & 	      & 	      & 	      & 	      & 	      & 	      & {\bf 801.707} & 	      \\
	      & 	      &               & 571.115       & 	      & 	      & 	      & 	      & 	      & 	      &               & {\bf 801.754} \\
	      & 	      & 571.541       & 	      & 	      & 	      & 	      & 	      & 	      & 	      & {\bf 801.787} & 	      \\
	      & 	      & 571.712       & 	      & 	      & 	      & 	      & 	      & 	      & 	      &               & {\bf 801.873} \\
	      & 	      &               & 572.101       & 	      & 	      & 	      & 	      & 	      & 	      &               & {\bf 801.913} \\
	      & 	      &               & 572.873       & 	      & 	      & 	      & 	      & 	      & 	      & {\bf 801.986} & 	      \\
	      & 	      & {\bf 574.444} & 	      & 	      & 	      & 	      & 	      & 	      & 	      &               & {\bf 802.033} \\
	      & 	      &               & {\bf 574.544} & 	      & 	      & 	      & 	      & 	      & 	      & {\bf 802.146} & 	      \\
	      & 	      &               & 574.886       & 	      & 	      & 	      & 	      & 	      & 	      &               & {\bf 802.153} \\
	      & 	      & 575.212       & 	      & 	      & 	      & 	      & 	      & 	      & 	      &               & {\bf 802.233} \\
	      & 	      &               & 576.171       & 	      & 	      & 	      & 	      & 	      & 	      & {\bf 802.305} & 	      \\
	      & 	      &               & {\bf 576.299} & 	      & 	      & 	      & 	      & 	      & 	      &               & {\bf 802.353} \\
	      & 	      & {\bf 576.450} & 	      & 	      & 	      & 	      & 	      & 	      & 	      & {\bf 802.385} & 	      \\
	      & 	      &               & {\bf 576.642} & 	      & 	      & 	      & 	      & 	      & 	      &               & {\bf 802.432} \\
	      & 	      &               & 576.770       & 	      & 	      & 	      & 	      & 	      & 	      & {\bf 802.465} & 	      \\
	      & 	      &               & 576.813       & 	      & 	      & 	      & 	      & 	      & 	      & {\bf 802.544} & 	      \\
	      & 	      & {\bf 577.473} & 	      & 	      & 	      & 	      & 	      & 	      & 	      & {\bf 802.584} & 	      \\
	      & 	      &               & {\bf 577.540} & 	      & 	      & 	      & 	      & 	      & 	      &               & {\bf 802.592} \\
	      & 	      & {\bf 578.070} & 	      & 	      & 	      & 	      & 	      & 	      & 	      & {\bf 802.624} & 	      \\
	      & 	      &               & {\bf 578.096} & 	      & 	      & 	      & 	      & 	      & 	      & {\bf 802.704} & 	      \\
	      & 	      & {\bf 578.240} & 	      & 	      & 	      & 	      & 	      & 	      & 	      & {\bf 802.744} & 	      \\
	      & 	      &               & {\bf 578.310} & 	      & 	      & 	      & 	      & 	      & 	      &               & {\bf 802.872} \\
	      & 	      & 579.519       & 	      & 	      & 	      & 	      & 	      & 	      & 	      &               & 803.151       \\
	      & 	      & 579.732       & 	      & 	      & 	      & 	      & 	      & 	      & 	      &               & 803.391       \\
	      & 	      &               & 581.517       & 	      & 	      & 	      & 	      & 	      & 	      &               & {\bf 803.990} \\
	      & 	      &               & 584.251       & 	      & 	      & 	      & 	      & 	      & 	      & {\bf 804.059} & 	      \\
	      & 	      &               & 584.550       & 	      & 	      & 	      & 	      & 	      & 	      &               & 804.349       \\
	      & 	      &               & 585.190       & 	      & 	      & 	      & 	      & 	      & 	      & 804.975       & 	      \\
	      & 	      &               & 585.275       & 	      & 	      & 	      & 	      & 	      & 	      & {\bf 805.015} & 	      \\
	      & 	      &               & 585.361       & 	      & 	      & 	      & 	      & 	      & 	      &               & {\bf 805.187} \\
	      & 	      &               & 585.659       & 	      & 	      & 	      & 	      & 	      & 	      & {\bf 805.254} & 	      \\
	      & 	      &               & 585.915       & 	      & 	      & 	      & 	      & 	      & 	      & {\bf 805.334} & 	      \\
	      & 	      &               & 586.299       & 	      & 	      & 	      & 	      & 	      & 	      & 805.453       & 	      \\
	      & 	      &               & 586.342       & 	      & 	      & 	      & 	      & 	      & 	      &               & 805.785       \\
	      & 	      &               & 586.427       & 	      & 	      & 	      & 	      & 	      & 	      & 806.091       & 	      \\
	      & 	      &               & 586.897       & 	      & 	      & 	      & 	      & 	      & 	      & 806.569       & 	      \\
	      & 	      &               & 587.963       & 	      & 	      & 	      & 	      & 	      & 	      & {\bf 806.688} & 	      \\
	      & 	      &               & 588.432       & 	      & 	      & 	      & 	      & 	      & 	      &               & {\bf 806.743} \\
	      & 	      &               & 592.309       & 	      & 	      & 	      & 	      & 	      & 	      &               & {\bf 806.783} \\
	      & 	      & 	      & 	      & 	      & 	      & 	      & 	      & 	      & 	      &               & 806.942       \\
	      & 	      & 	      & 	      & 	      & 	      & 	      & 	      & 	      & 	      &               & 807.102       \\
	      & 	      & 	      & 	      & 	      & 	      & 	      & 	      & 	      & 	      &               & 807.261       \\
	      & 	      & 	      & 	      & 	      & 	      & 	      & 	      & 	      & 	      & {\bf 807.524} & 	      \\
	      & 	      & 	      & 	      & 	      & 	      & 	      & 	      & 	      & 	      &               & {\bf 807.580} \\
	      & 	      & 	      & 	      & 	      & 	      & 	      & 	      & 	      & 	      &               & {\bf 807.700} \\
	      & 	      & 	      & 	      & 	      & 	      & 	      & 	      & 	      & 	      &               & 807.860       \\
	      & 	      & 	      & 	      & 	      & 	      & 	      & 	      & 	      & 	      & 809.116       & 	      \\
	      & 	      & 	      & 	      & 	      & 	      & 	      & 	      & 	      & 	      &               & {\bf 810.052} \\
	      & 	      & 	      & 	      & 	      & 	      & 	      & 	      & 	      & 	      & {\bf 810.111} & 	      \\
	      & 	      & 	      & 	      & 	      & 	      & 	      & 	      & 	      & 	      &               & 811.008       \\
	      & 	      & 	      & 	      & 	      & 	      & 	      & 	      & 	      & 	      &               & 811.367       \\
	      & 	      & 	      & 	      & 	      & 	      & 	      & 	      & 	      & 	      & 811.861       & 	      \\
	      & 	      & 	      & 	      & 	      & 	      & 	      & 	      & 	      & 	      &               & 814.034       \\
	      & 	      & 	      & 	      & 	      & 	      & 	      & 	      & 	      & 	      & 814.444       & 	      \\
	      & 	      & 	      & 	      & 	      & 	      & 	      & 	      & 	      & 	      & 816.192       & 	      \\
	      & 	      & 	      & 	      & 	      & 	      & 	      & 	      & 	      & 	      & 816.232       & 	      \\
	      & 	      & 	      & 	      & 	      & 	      & 	      & 	      & 	      & 	      &               & 816.421       \\
	      & 	      & 	      & 	      & 	      & 	      & 	      & 	      & 	      & 	      &               & 817.455       \\
	      & 	      & 	      & 	      & 	      & 	      & 	      & 	      & 	      & 	      &               & 817.813       \\
	      & 	      & 	      & 	      & 	      & 	      & 	      & 	      & 	      & 	      &               & 819.363       \\
	      & 	      & 	      & 	      & 	      & 	      & 	      & 	      & 	      & 	      &               & 820.833       \\
	      & 	      & 	      & 	      & 	      & 	      & 	      & 	      & 	      & 	      &               & 822.978       \\
	      & 	      & 	      & 	      & 	      & 	      & 	      & 	      & 	      & 	      &               & 828.253       \\
	      & 	      & 	      & 	      & 	      & 	      & 	      & 	      & 	      & 	      &               & 828.293       \\
\tableline
%\end{tabular}
\end{longtable}
}
 %label={tab:lam_id}

We used the methods of  Paper 1 to detect wavelengths with significantly over-deep
transits. In Paper 1, we used  the average of orbits 1 and 5 for the out of transit (OOT) spectrum,
and orbit 3 for the in transit (IN) spectrum. As Fig.~\ref{lc_norm-v1} and Fig.~\ref{fig:1visfits}
demonstrates, there is no reason
to change this, so we used the same portions of data (re-reduced as detailed in Section~\ref{hstobs}).
Visit 2 shows no clear pre-ingress OOT data, so in this case we used orbit 5 for the OOT spectrum, and orbit
4 for the IN spectrum. Consequently the propagated error estimate for our Visit 2 OOT spectrum is larger
that that for the Visit 1 OOT spectrum.

%-wavelength identification-------------------------------------------------------------------
\begin{figure}
\begin{center}
\includegraphics[width=\hsize,clip]{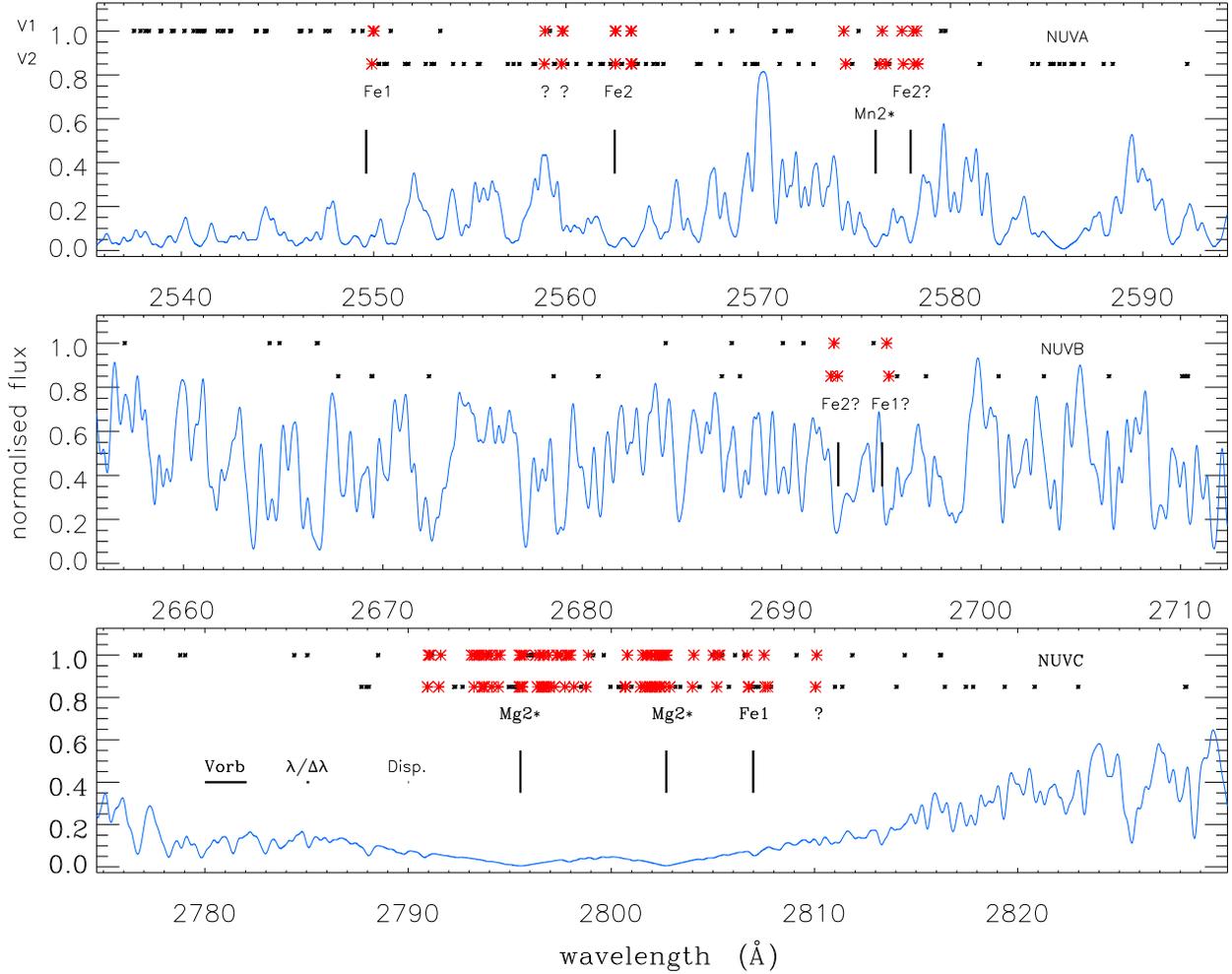}
% currently only have a pdf version
\caption{The wavelengths at which significantly over-deep transits were detected are indicated
 with a black cross, except for those wavelengths detected in both visits where
 we use a large red asterix instead. The upper row and lower rows of symbols correspond to Visit 1 and Visit 2
 respectively. The horizontal scale bars in the bottom (NUVC) panel
 indicate the spectral dispersion, the spectral resolution, and the magnitude of WASP-12\,b's orbital
 velocity. The blue line is the synthetic stellar spectrum, normalised to the continuum level, with a resolution of $R=10,000$, adopted to simulate the
 HST LSF. We do not use the synthetic stellar spectrum in our analysis. Wavelengths of a few absorbing species
 are identified, with a small asterix after the ion identification for resonance lines.
}
\label{lam_id}
\end{center}
\end{figure}
%--------------------------------------------------------------------

For each visit, we define the ratio spectrum, $d_i$  (= IN / OOT ), where IN is the in-transit flux and OOT is
the out of transit flux, to measure the transit depth at wavelength $\lambda_i$.
We examine $d_i$ to determine wavelengths $\lambda_i$ with statistically significant excess transit depths.
These are the wavelengths where the diffuse gas absorbs. These individual
contributions combine to produce the excess depth in the
 light curves examined in \S\ref{sec:nuv_lc}.
We report the wavelengths of these points in Table~\ref{tab:lam_id}
and in Fig.~\ref{lam_id}. In Table~\ref{tab:lam_id} we use two different criteria to assess anomalously
low points, i.e. those deviating by 3$\sigma$, as we did in Paper 1. The first of these examines the ratio spectrum against the error estimate we
obtain from the propagated (Poisson-dominated) errors in the reduction. We label this error estimate as $\sigma{d_{i}|_{prop}}$, and report in Table~\ref{tab:lam_id} wavelengths where the ratio spectrum is 3$\sigma{d_{i}|_{prop}}$ or more below the median value of the ratio. These wavelengths are also indicated
in  Fig.~\ref{lam_id}. The second assessment uses the RMS scatter, $\sigma_{d_{i}|_{exp}}$, within the ratio spectrum itself in the criterion. As Fig. 3 of Paper 1 showed, the  $\sigma_{d_{i}|_{exp}}$ criterion is more likely to pick out
points where the signal in the observed stellar spectrum is low.

Our ratio spectra have a total of just under 3100 points, so if the noise is Gaussian-distributed we expect
$\sim 9$ distinct points in each to deviate by more than 3$\sigma{d_{i}|_{prop}}$. As Table~\ref{tab:lam_id} shows, we find
many more deviating points than this. In every case the points which deviate by more than 3$\sigma{d_{i}|_{prop}}$
are deviations below the median of the ratio spectrum, indicating extra absorption compared to the median. Where there
are positive deviations, these are accompanied by large values of $\sigma{d_{i}|_{prop}}$ due to the low flux, and hence low SNR
of the WASP-12 spectrum at that wavelength. Consequently, none of these positive deviations are 3$\sigma{d_{i}|_{prop}}$.
We would expect, nonetheless, to find a small number ($\sim 4 \pm 2$) of positive deviations due to white noise,
but we do not. Possibly our propagated errors are slightly over-estimated. This leads us to conclude
we have detected statistically significant excess absorption at $\sim 200$  distinct wavelengths in each visit
from the gas around WASP-12.

As noted in \S\ref{sec:nuv_lc} the NUVA and NUVC spectral regions are more strongly absorbed
within the stellar photosphere than is NUVB. In \S\ref{sec:nuv_lc} we showed that the
transit is deeper in wavelength regions where the photospheric absorption is strong. These
wavelength regions contain overlapping spectral lines which produce strong photospheric absorption
of the underlying stellar continuum. To the extent that the extended diffuse gas cloud has similar abundances
and temperature as the stellar photosphere, it too will produce strong absorption
in these wavelength regions.
Fig.~\ref{lam_id} supports this: it is immediately obvious that
there are far more wavelengths exhibiting enhanced transit depths in NUVA and NUVC than there are
in NUVB. It is also clear that the wavelengths exhibiting excess transit depths generally occur where the emergent
stellar flux is low. This shows our rule of thumb works well: the
photospheric absorption within a wavelength range is a good  predictor of the likely diffuse absorption.

Fig.~\ref{lam_id} shows that our identified wavelengths in NUVC from the two distinct visits
are largely associated with absorption in the wings of the \ion{Mg}{2} resonance lines. This is clearly and unambiguously
detected in both visits, and at so many distinct wavelength-pixels that the median value of the ratio spectrum
must be lowered by them. Away from the cores of \ion{Mg}{2} at $\sim 2788 - 2789 \AA$, $\sim 2811 - 2812 \AA$ and $\sim 2815\AA$ there is a suggestion of a consistent wavelength shift between the Visit 2 (upper) and Visit 1 (lower) detections, with the Visit 1 transits being redshifted by $\sim 0.5 - 1 \AA$ ($\sim$ 50-100\,\kms). This shift is much greater than our
spectral resolution, but less than the magnitude of the planet's orbital velocity (230 \, \kms), which is the natural velocity
scale for motions of material orbiting the star.
The NUVA and NUVB identified wavelengths show less consistency between the two visits, except for wavelengths consistent with resonance lines which are picked out in both visits.

The next step is to attempt to identify the spectral lines causing these enhanced transit
depths.
In Paper I we arbitrarily restricted our line identifications to resonance lines, and the consistently detected
wavelengths in Fig.~\ref{lam_id} show this was a sensible first step.
Nonetheless, strong non-resonance lines of abundant elements may be more prominent
than the resonance lines of rare elements. We attempted to assess this with
a procedure which accounts for the stellar abundances derived for WASP-12 \citep{fossati2010b};
the velocity shift required to match the rest wavelength of the line to the observed
deviation; the excitation potential of the line; and the ${gf}$ value of the line.
We introduce three simple functions: $P_{a}$, $P_{\Delta\lambda}$, and $P_{\chi}$, (each defined below) to estimate the effects of
the first three of these
factors on the  likelihood of a particular identification being correct.

We wish to estimate the probability that an observed
3$\sigma$ deviation at wavelength $\lambda$,
arises from absorption by
spectral line $i$ of element $el$ which has rest wavelength $\lambda _i$.
We consider a wavelength range $\lambda _i \pm \Delta\lambda$ and adopt $\Delta\lambda=$
3\,\AA\ in our analysis to recover the full range of
deviating points we identified in the broad wings of the
\ion{Mg}{2} lines in Paper 1 (see Fig.~3 therein.) This corresponds to a velocity
shift slightly greater than the magnitude of the orbital velocity of WASP-12\,b, but our procedure
favours the smallest possible value of wavelength/velocity shift.

The wavelength shift required to match the observed point to rest wavelength
is accounted for in the factor $P_{\Delta\lambda}$; the excitation potential, $\chi_i$,
of the lower level in the factor $P_{\chi}$; and the abundance in the factor
$P_{a}$. We define these probability factors by
\begin{equation}
P_{\Delta\lambda}=1-{\left ( {{|\lambda _i-\lambda|}\over{\Delta\lambda}} \right )}; \quad ~~~~~~~
% P_{\chi}=e^{-\chi_i}? \quad ~~~~~~~~
P_{\chi}=e^{-\frac{\chi_i}{kT}} \quad ~~~~~
% P_a=\frac{10^{\log (N_{el}/N_{\rm tot})}}{Z}? \quad ~~~~~~
P_a=\frac{N_{el}}{Z}
\end{equation}
where $N_{el}$ is the abundance of the element considered, and $Z$ is the
`metal' abundance.
Each gives a value of unity
for the most favorable value of the quantity concerned;
$P_{a}$ is unity if the proposed  gas composition comprises only H, He and the
element under consideration.
$P_{\Delta\lambda}$ was designed to have a weak dependence on the shift $|\lambda _i-\lambda|$
as we expect the gas to have non-zero velocity w.r.t. the stellar rest-frame.
$P_{\chi}$ is proportional to the expected population in the lower level, and $P_{a}$
is proportional to the number of absorbing atoms/ions.

$P_{\Delta\lambda}$, $P_{\chi}$ and $P_{a}$   are multiplied together and further multiplied
by the ${\it gf}$ value of the line under consideration to give our estimated overall probability measure, $P_i$,
that the deviation is associated with the specified spectral line, $i$. P$_i$ in Eq.\ref{eqn:pdef} is thus not
normalised to 1.
\begin{equation}\label{eqn:pdef}
P_i=(P_{\Delta\lambda} P_{\chi} P_{a}) \cdot ({\it gf})_i
%\cdot 10^{log {\rm gf}\,_i}
\end{equation}

We then use this framework to assess the likely contribution to the measured deviation of all known spectral lines within
$\Delta\lambda$. $P_i$ estimates the likely relative contribution
of each of these lines to the detected absorption;  summing over all lines
estimates the total absorbing capacity, $P_{\rm tot}$. The fraction of this
due to any given spectral line is our proxy for the likelihood, $L_i$ that
this is the correct line identification.
\begin{equation}
P_{\rm tot}=\sum_{i}P_i ; \quad L_i = \frac{P_i}{P_{\rm Tot}}
\end{equation}

\begin{figure}
\begin{center}
\includegraphics[width=\hsize,clip]{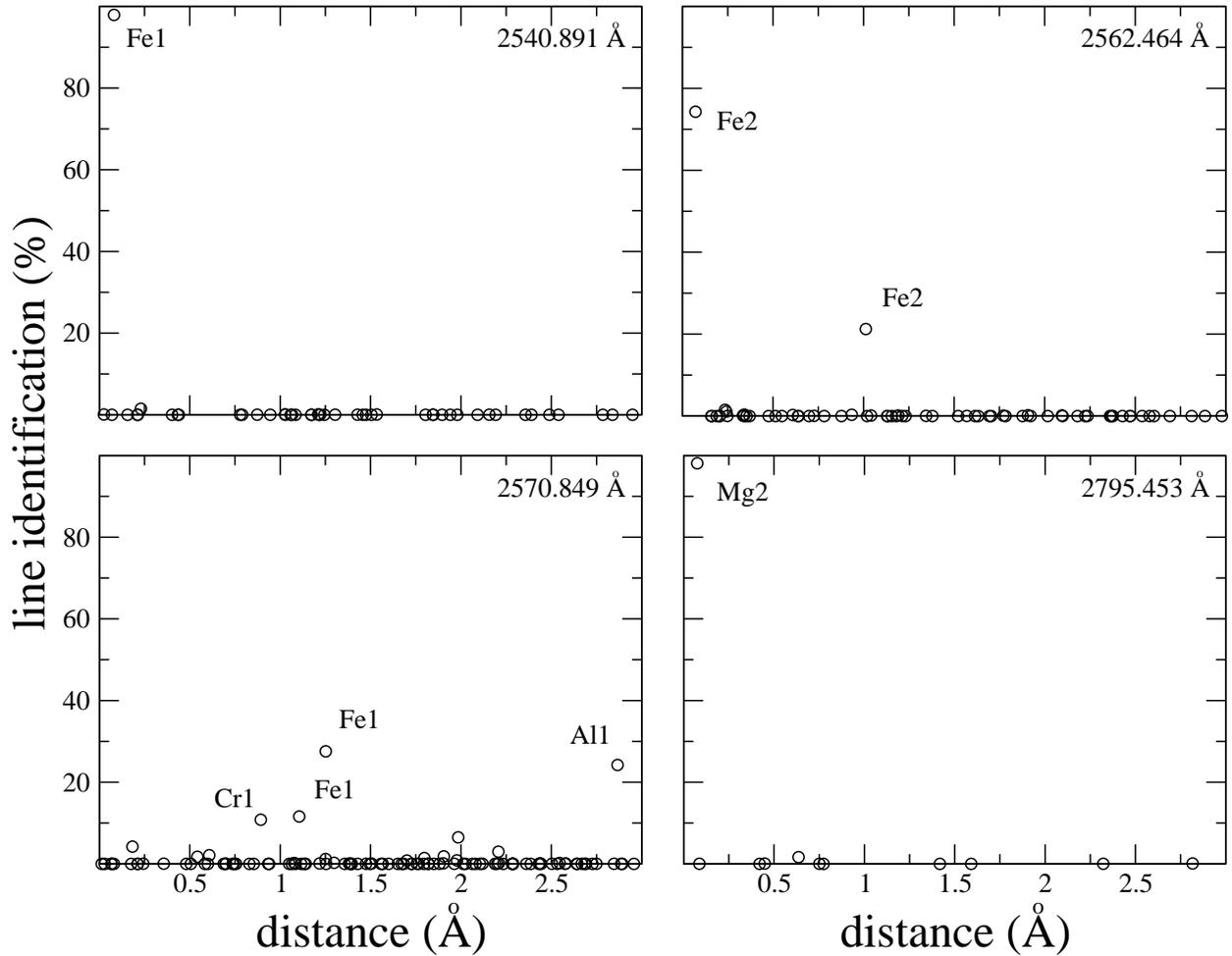}
\caption{Line identifications from the Visit 1 data. The adopted temperature
was 3000\,K and an abundance pattern consistent with \citet{fossati2010b} was assumed.
The examples show three fairly firm identifications and one point where no firm conclusion is
possible.
$L_{i}$ is shown for every spectral line where it exceeded a value of $10^{-5}$, many
lines considered had values of $L_i$ below this threshold.}
\label{line_id_v1}
\end{center}
\end{figure}

\begin{figure}
\begin{center}
\includegraphics[width=\hsize,clip]{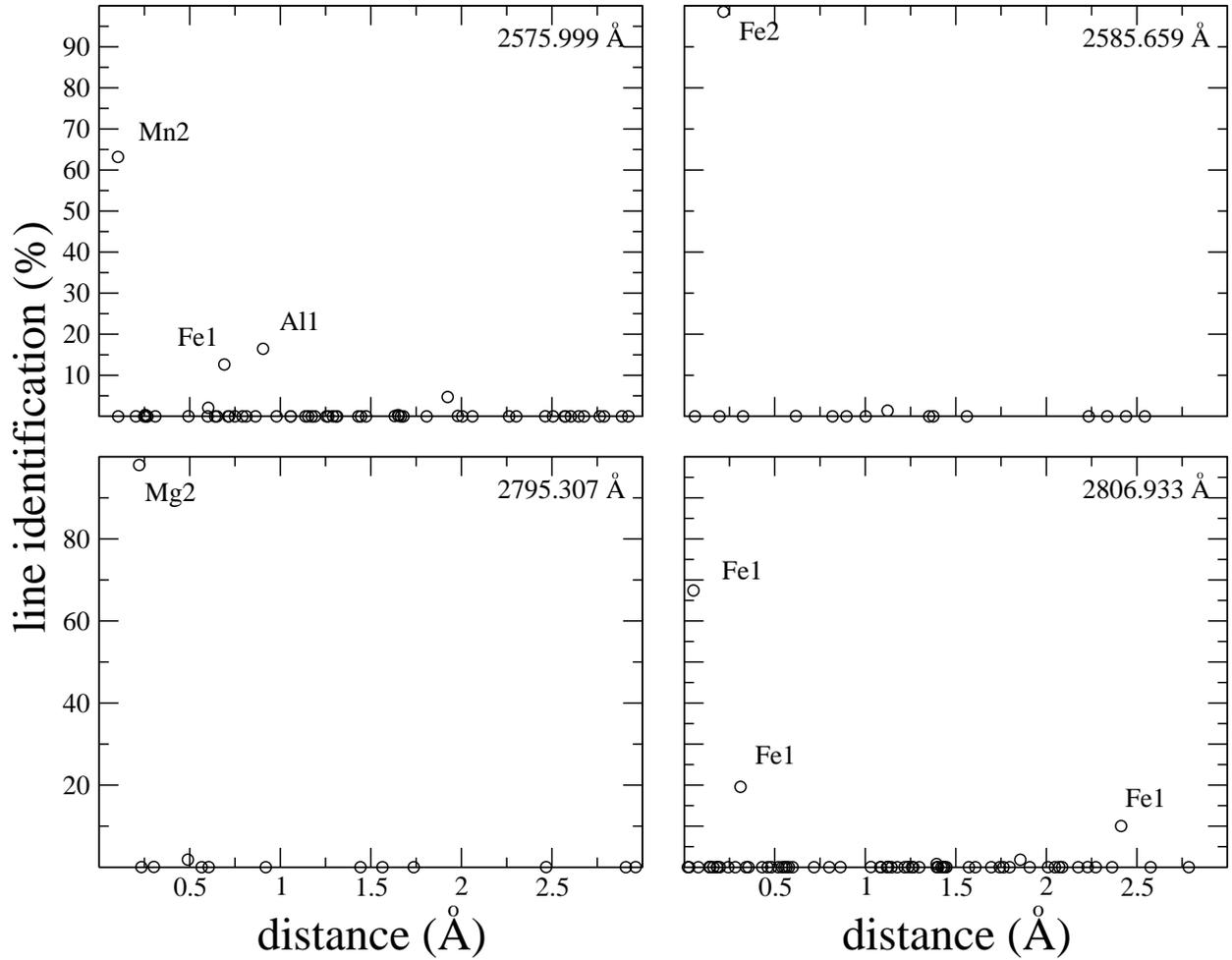}
\caption{Line identifications from the Visit 2 data. The adopted temperature
was 3000\,K and an abundance pattern consistent with \citet{fossati2010b} was assumed.
The examples show three fairly firm identifications and one (top left) probable identification.
$L_{i}$ is shown for every spectral line where it exceeded a value of $10^{-5}$, many
lines considered had values of $L_i$ below this threshold.}
\label{line_id_v2}
\end{center}
\end{figure}

This framework indicates, with some physical justification for our
 estimates, how the likely line identification depends on the assumed
temperature and abundance pattern. Our factor to account for a decreasing likelyhood
as the shift from the line's rest wavelength increases is arbitrary, but
adequate, given the complex COS LSF, the unknown velocity distribution in
the absorbing gas, and the SNR of our data. The temperature dependence is only
indicative: we have not used the Saha equation to determine the ionic balance. Figure~\ref{line_id_v1} shows
examples of applying this procedure to a selection of four points detected as
having enhanced transit depths at significance of  $3\sigma$ or more in Visit 1.
Similarly Figure~\ref{line_id_v2} shows a selection of line ID assessments
from the Visit 2 data.

%-line identification-------------------------------------------------------------------
%--------------------------------------------------------------------
%--------------------------------------------------------------------
%--------------------------------------------------------------------
In Fig.~\ref{line_id_v1} the two upper panels reveal
absorption from \ion{Fe}{1}  and \ion{Fe}{2} causing enhanced transit depths at  2540.891~$\AA$ and 2562.464~$\AA$.
The lower right panel shows the overwhelming probability that the
enhanced transit depth at 2795.453~$\AA$ is caused by
\ion{Mg}{2}. This is because the imputed line is one of the pair of resonance lines
with large ${\it gf}$ values which dominate the NUVC spectral region. For this line the $P_{\chi}$
and ${\it gf}$ factors are so large that the assumed \ion{Mg}{2} abundance would need to be
vanishingly small to identify any other nearby spectral line as likely to contribute significantly
to the total absorption.
Our Paper 1 detection of absorption in  \ion{Mg}{2} during Visit 1
remains unassailable under
this more careful scrutiny.
Visit 2 provides independent data, and as the lower left hand panel
of Fig.~\ref{line_id_v2} demonstrates, it shows the Visit 2 detection of an enhanced transit depth
at 2795.307~$\AA$
can be confidently attributed to \ion{Mg}{2}.
Note the wavelength solutions produced by the simultaneous arc lamp spectra differ for the two visits,
so we cannot compare pixels with identical wavelength sampling without rebinning. Rebinning is undesirable because
adjacent pixels are then no longer independent.

Not all deviating points could be unambiguously associated with absorption
by a single ionic species: the lower left panel of Fig.~\ref{line_id_v1}
and the upper left panel of Fig.~\ref{line_id_v2}
show examples where there are plausibly large contributions from a number of different elements.
The detected absorption is likely to be a mix
of the various spectral lines, with the dominant contributor(s) depending on the
the unknown composition and physical conditions of the gas.
We can, however, conclude from Figs.~\ref{line_id_v1} and ~\ref{line_id_v2}
that we have detected absorption  from \ion{Fe}{1}, \ion{Fe}{2}, \ion{Mg}{2},
and probably \ion{Mn}{2}.

Table~\ref{tab:lam_id} gives the full list of wavelengths where we detected
enhanced transit depths. We used a relative shift of $0.2 \AA$ as
our measure of wavelength consistency, this is approximately the value of the
formal resolution plus the dispersion (the wavelength solution and hence
the pixelation of the ratio spectra differs between the two spectra). It is
{\it far} less than the magnitude of WASP-12\,b's orbital velocity, so as the
motions in the gas are probably time-dependent, we are imposing a very strict
criterion for consistency between the two visits. Where enhanced transit depths
are detected within $0.2 \AA$ of each other in both visits, these are listed
in bold in Table~\ref{tab:lam_id}.  We have two independent measurements of the ratio spectrum,
one from each visit, so enhanced depths which are detected at consistent wavelengths
in both are unlikely to be due to noise. We would expect a match for any wavelength pixel
picked out in Visit 1 to occur due to gaussian-distributed noise in 2.7\% of cases.
In NUVA (NUVB, NUVC) 53\% (28\%, 70\%) of Visit 1 detections using the 3$\sigma{d_{i}|_{prop}}$
statistic are matched by the same statistic in Visit 2.

\begin{figure}
\begin{center}
\includegraphics[width=130mm]{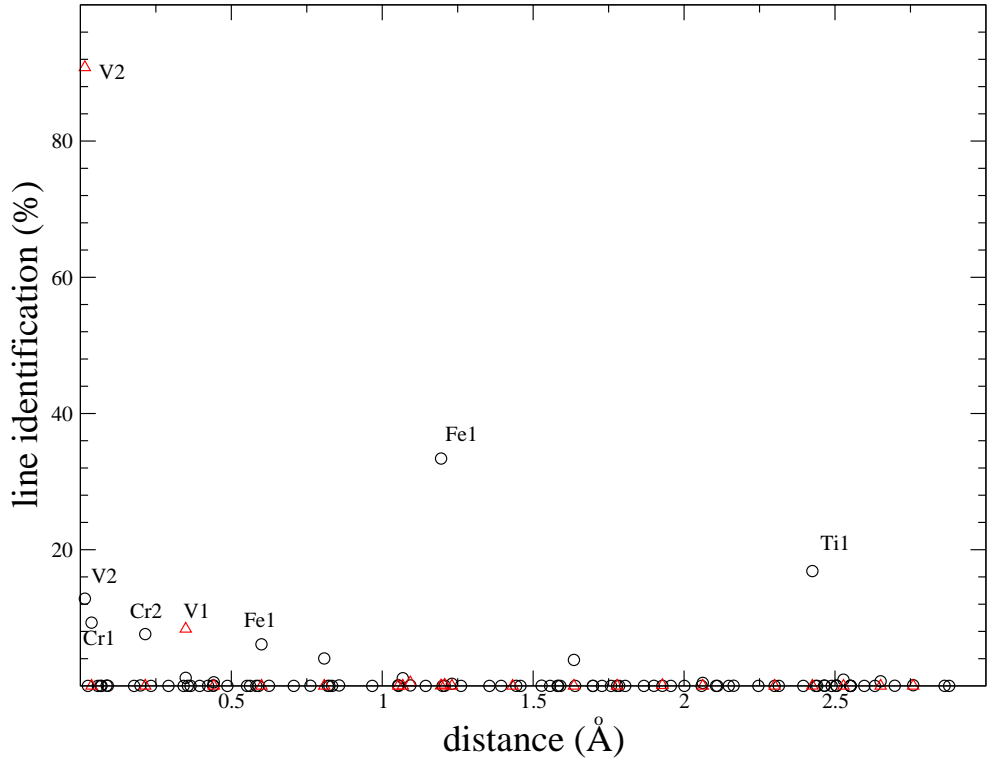}
\caption{ Results of applying our line identification procedure to the deviating point
at $2672.018$\,\AA. The black points assume our best estimate of the abundances of the WASP-12
stellar photosphere, while the red triangles instead apply the analysis with an arbitrary
increase in the abundance of vanadium so that it has the same abundance as iron (by number).
This makes it clear that the results of our line identification are strongly dependent
on the assumptions we make.
}
\label{fig:v_or_not_v}
\end{center}
\end{figure}

A definitive assignment of identifications to these many wavelengths exhibiting
enhanced transit depths requires more certain knowledge of the properties of the
diffuse gas than we currently have. We can, however, examine how changing our
assumptions affects our line identifications.
Fig.~\ref{fig:v_or_not_v} examines how the identification of the absorber at $2672.018 \AA$
depends on the assumed abundance distribution.
In Paper 1, by including only resonance lines in our analysis, we excluded many
plausible spectral lines from our consideration. This led us to identify enhanced
absorption at wavelengths
around 2680\, $\AA$ with \ion{V}{2} (see Table 2 of Paper 1). Fig.~\ref{fig:v_or_not_v}
shows a more careful examination of the line identification of one of these
points.
The black open circles show our likelihood estimates for an abundance pattern
which matches our current best estimate
of that of the WASP-12 stellar photosphere \citep{fossati2010b}. With this abundance pattern,
vanadium is too rare to make a dominant contribution to the absorption; instead we are
led to conclude that \ion{Fe}{1} and \ion{Ti}{1} are more likely than \ion{V}{2}.
Iron is generally rather abundant, and in common with many other iron-peak elements including vanadium, has a complex electronic structure, leading to a large
number of blended iron lines, particularly in the NUV.
If, on the other hand, we were to assume that vanadium is abundant in WASP-12\,b's exosphere
and we arbitrarily increased the vanadium abundance to match that of iron in WASP-12's
photosphere, we would introduce a significant change to our assessment of the line identification.
The open red triangles in Fig.~\ref{fig:v_or_not_v} show the results
we obtain if we adopt this assumption. In this case, we identify \ion{V}{2} and \ion{V}{1} as the
species most likely to be responsible for the enhanced transit depth. It is worth noting that VO has been
widely discussed as a possible constituent of the stratospheres of hot Jupiter exoplanets  \citep[e.g.][]{Desert08,Speigel09, Fortney10} and it is possible that the gases lost from the upper atmosphere of a hot Jupiter could have
enhanced vanadium abundances.

%--------------------------------------------------------------------
%--------------------------------------------------------------------

%--------------------------------------------------------------------
\begin{figure}
\begin{center}
\includegraphics[width=130mm]{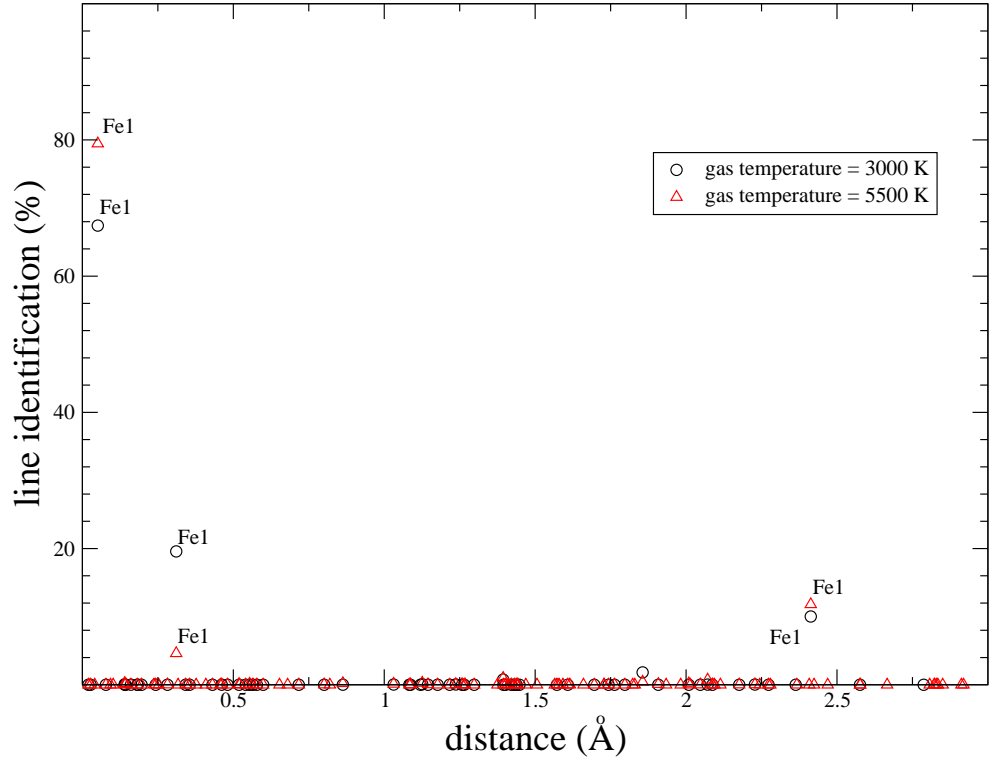}
\caption{Results of applying our line identification procedure to the deviating point
at $2806.933-T5500$\,\AA. The two sets of points correspond to 3000\,K and 5500\,K.
 This makes it clear that the results of our line identifications depend on the
 temperature as well as the abundance pattern assumed for the absorbing gas.
}
\label{fig:line_id_Tdep}
\end{center}
\end{figure}
%--------------------------------------------------------------------

Fig.~\ref{fig:line_id_Tdep} shows an example
of how the imputed line differs for values of 3000\,K and 5500\,K in $P_{\chi}$, both plausible values for
the exosphere of WASP-12.
None of our clear identifications changed between these two temperatures, but we note
we have not included the temperature dependence of the ionisation balance in our assessments.

Figs.~\ref{fig:v_or_not_v} and~\ref{fig:line_id_Tdep} show that
our line identifications depend on the assumptions we make.
 This is generally true in astrophysics,
but in more mature fields, we have a sound basis for confidently adopting likely assumptions. For
example we have a very good understanding of the physical conditions prevailing in stellar atmospheres,
and much high SNR ratio data has been used to hone models such as the one we plotted in Figs.~\ref{flux-v1ev2} and~\ref{lam_id}.
The study of hot Jupiter exospheres is not so well-developed.

To be sure of all line identifications in WASP-12\,b and other
hot Jupiters, we need sound assumptions for
the likely abundances in the absorbing material, and measurements
of its physical properties. Hot Jupiter atmospheres probably
have prodigious winds and disequilibrium chemistry which will affect
the abundances of the exospheric material  \citep[e.g.][]{Moses11}.
The models have many degrees of freedom, and
for WASP-12\,b  constraining them will be challenging, as
the system is relatively distant  \citep[380$\pm$85\,pc][]{fossati2010b}
and hence at the limit of what we can do with HST/COS.
WASP-12\,b is roughly the most distant exoplanetary system for which
NUV transit observations can realistically be obtained with technology available in
the forseeable future.
The best chance to put our understanding of hot Jupiter exospheres
on a firm foundation would be to observe
the two brightest hot Jupiters, HD\,209458\,b and HD\,189733\,b
in the NUV.
In some cases, however, we already have
robust identifications of spectral lines exhibiting
enhanced transit depths in WASP-12\,b, for example
for the \ion{Mg}{2} resonance lines.

\subsection{Chromospheric Activity in WASP-12 and Tenuous Gas Surrounding the System}
\label{sec:act}
%--------------------------------------------------------------------
\begin{figure}
\begin{center}
\includegraphics[width=\hsize,clip]{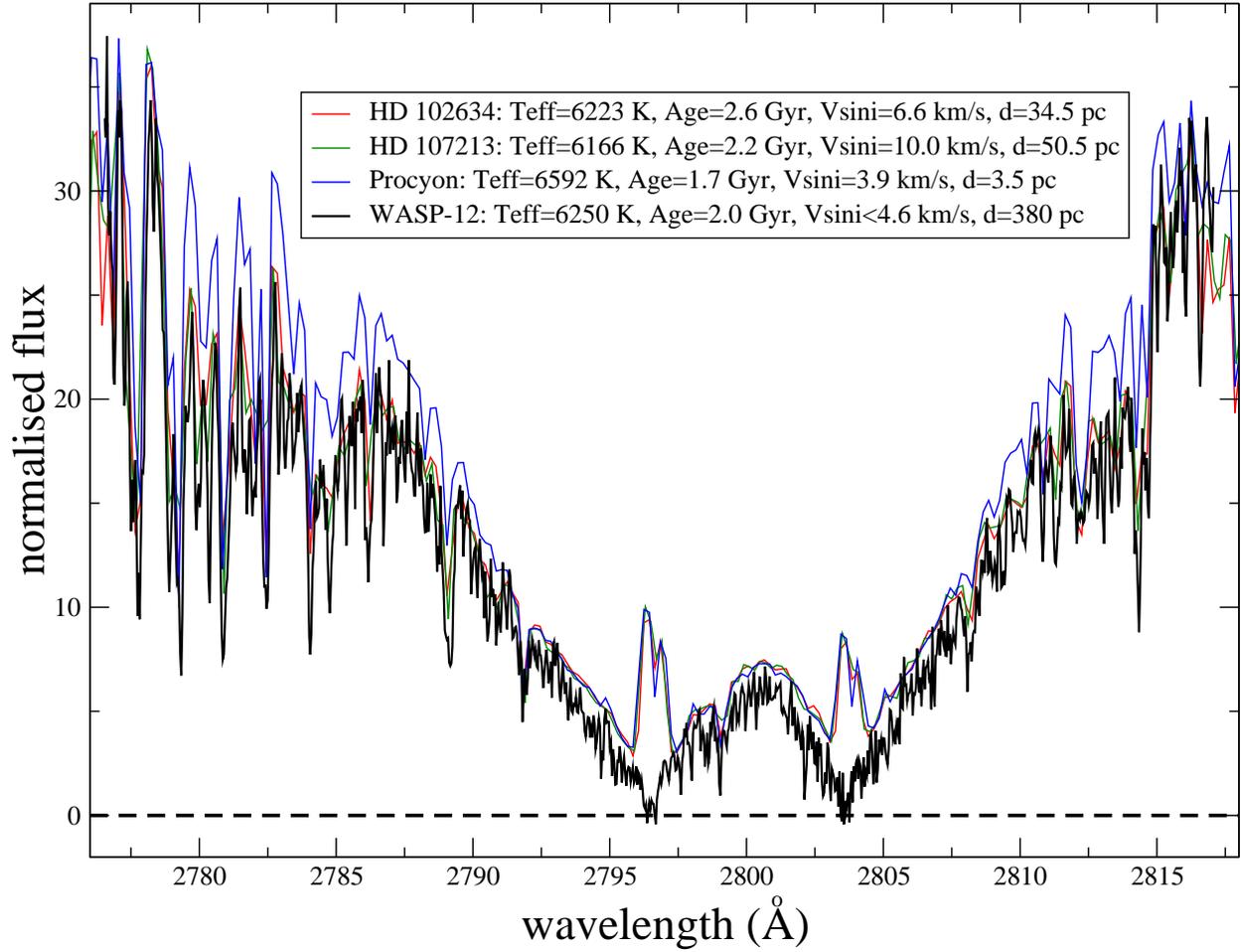}
\caption{Our Visit 1 NUVC coverage of the cores of the \ion{Mg}{2} resonance lines compared with those
of three other stars (see text). The
cores of the WASP-12 lines show no emission whatsoever, and are suggestive of narrow absorption components
which reduce the flux in the cores of these lines to {\it zero}. WASP-12 is remarkable in
the appearance of its \ion{Mg}{2} resonance line cores.
}
\label{fig:mgii_core}
\end{center}
\end{figure}
%--------------------------------------------------------------------

Our HST/COS data raise questions concerning stellar activity in WASP-12.
Can we attribute the high point at phase $\phi=0.95$ in the Visit 2 light curves
to stellar activity as we speculated in \S\ref{sec:nuv_lc}? \citet{fossati2010b}
found an age of 1.0-2.65 Gyr for WASP-12, and an effective temperature of
$T_{\rm eff} = 6250 \pm 100 \, {\rm K}$, in agreement with \citet{hebb2009}
but with tighter uncertainties.
Generally, stars of this \Teff\ and age are expected to exhibit activity, and
emission cores in the MgII resonance lines are one clear observable consequence of this.
As Figure~\ref{fig:mgii_core} demonstrates, however, MgII emission cores are
conspicuously absent in WASP-12.

To produce Figure~\ref{fig:mgii_core} we normalised the spectra of WASP-12 and two stars
of the same age and
\Teff\ to match the flux in the far wings of the Mg II lines (the regions used in the normalisation are
further from the line cores than the edges of Fig.~\ref{fig:mgii_core}).
These two stars, HD\,102634 and HD\,107213, are plotted in red and green in Fig.~\ref{fig:mgii_core}.
As expected for such similar stars, the line profiles of WASP-12, HD\,102634 and HD\,107213
match well throughout the profile of the \ion{Mg}{2} lines, except within about $10\,\AA$ of either
of the line cores, where WASP-12's spectrum lies below the others. Fig.~\ref{fig:mgii_core} also
shows Procyon, which is a commonly used reference star, slightly hotter, slightly younger and
more slowly rotating \citep[$v\sin i = 3.9\,{\rm km\,s}^{-1}$,][]{Schroder09} than
the other three objects. Procyon is noted for its low stellar activity.
Procyon, HD\,102634 and HD\,107213 all have very similar line profiles between $2792\,\AA$ and $2807\,\AA$,
with similar  emission cores in all three cases surrounded by fairly steep declines in
flux towards the line cores, with very similar gradients in all three examples. WASP-12
has anomalously low fluxes within about $10\,\AA$ of the \ion{Mg}{2} line cores, with
steeper
gradients towards the line cores and absolutely no sign of emission in the line cores.

Figure~\ref{fig:mgii_core} shows that WASP-12, HD\,102634 and HD\,107312, which have similar
ages and temperatures, match extremely well, except for in the line cores.
Indeed, the WASP-12 line cores show no sign of the anticipated
emission reversals, and appear essentially as saturated
absorption profiles:
the line cores have {\it zero} flux.
As the point-to-point deviations in the WASP-12
spectrum make clear, we would definitely have detected
emission cores in WASP-12 if even if they were present at a strength significantly less than
that in the three comparison stars.
Even if WASP-12 were a slowly-rotating sub-giant, \ion{Mg}{2} emission cores would be expected \citep{2010MmSAI..81..553A}. Even very inactive Sun-like stars have significant
chromospheric spectra with prominent \ion{Mg}{2} emission powered by acoustic shock heating
and local dynamo action which converts hydrodynamic turbulence into
magnetic flux.

Our interpretation of Fig.~\ref{fig:mgii_core} is that the inner parts of WASP-12's Mg II lines
are absorbed by gas beyond the stellar chromosphere, with sufficient
column density to completely obliterate the expected emission reversals,
and to perhaps also depress the line wings just outside the cores.  Because the
star is distant, one possibility is interstellar absorption.
As we demonstrate below, however, the necessary Mg$^{+}$ column density
is quite substantial, and is implausible unless the ISM in that
direction is unusual.  Another possibility, which we favor, is that the absorption
is local to the WASP-12 system.

Gas immediately surrounding WASP-12\,b cannot account for the absorption we have just
described. The planet itself covers only a small fraction of the stellar disc: to completely
remove
the predicted Mg II chromospheric emission would require a much more spatially
extended distribution of absorbing material, blanketing the entire stellar disc.
We propose, in fact, that the entire system is
shrouded in diffuse
gas, very likely stripped from WASP-12\,b itself under the harsh radiation and stellar wind
conditions so close to the parent star.
\citet{2010ApJ...720.1569K} measured the \ion{Ca}{2}~H \& K emission lines of 50
transiting planet host stars including WASP-12. WASP-12's \ion{Ca}{2}~H \& K lines
are completely devoid of detected emission cores, as are the majority of \citet{2010ApJ...720.1569K}'s
sample. \citet{2010ApJ...720.1569K} interpreted their results in terms of a correlation
between stellar activity and hot Jupiter atmosphere type, but
our hypothesis suggests an alternative/additional explanation. Many
of these systems could be shrouded in diffuse gas which absorbs any emission cores
produced by stellar activity.

Chromospheric activity is strongly correlated with stellar age and rotational velocity. WASP-12's
rotational velocity is unknown, but the transit of WASP-12\,b is accompanied by an undetected
Rossiter-McLaughlin (RM) effect (Husnoo et al. 2011), so either the rotational velocity is low or
the orbital angular momentum of WASP-12 b is misaligned with the stellar rotation axis (see e.g.
Haswell 2010, for explanation).  If the chromospheric activity is low, as expected for a middle-aged
slowly rotating dwarf, a stellar flare is
a very unlikely
explanation for the high flux in orbit 3 of our Visit 2 light curves. We must then conclude
either our data are intrinsically noisy beyond the assigned photometric error,
or that we are viewing WASP-12 through diffuse gas
at all observed orbital phases, and there happened to be a relatively clear view to the stellar
surface during orbit 3 of Visit 2. This latter possibility might be produced by a bow shock
in a very extended planetary magnetosphere \citep{vidotto11}.

\subsection{Resonance Line Absorption in WASP-12 and the ISM}
\label{sec:abs}

%--------------------------------------------------------------------
\begin{figure}
\begin{center}
\includegraphics[width=90mm]{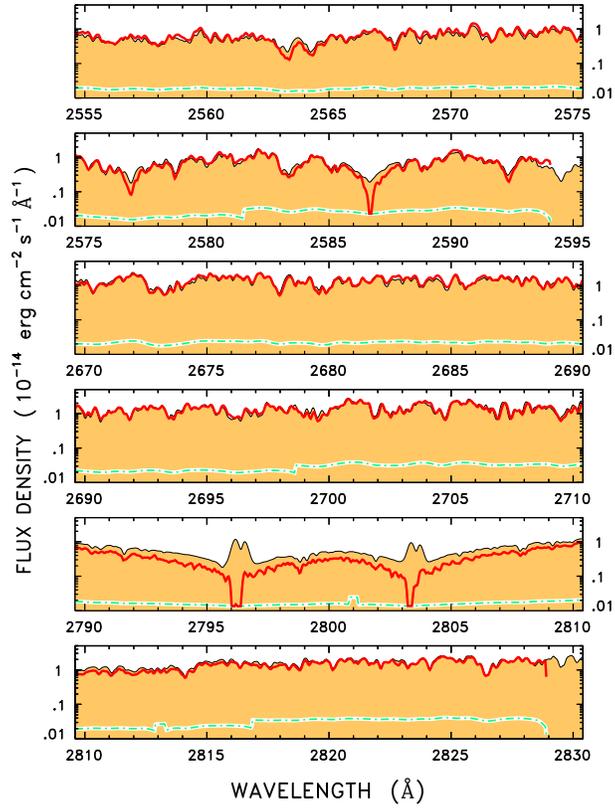}
\caption{Our WASP-12 NUV spectrum (red) in comparison to that of $\alpha$\,Cen (black line, filled beneath). A logarithmic flux scale has been adopted to emphasize the deepest absorption. The green dotted line
indicates the errors on the WASP-12 spectrum: data are truncated at levels consistent with
zero flux.
}
\label{fig:UVres_W12_alphcen}
\end{center}
\end{figure}
%--------------------------------------------------------------------

Figure~\ref{fig:UVres_W12_alphcen}  shows our Visit 2 WASP-12 spectrum compared to HST/STIS
data on  $\alpha$\,Cen from the archive. The STIS data have been convolved with the COS NUV LSF
and scaled by the multiplicative factor $ 4 \times 10^{-5}$.
Both spectra have been smoothed for display purposes.
The flux axis is logarithmic to emphasize
the deep absorption in the cores of strong lines.
The photometric
error is per resolution element, and the flux densities have been truncated at
the $1\,\sigma$ error level.
As in Fig.~\ref{fig:mgii_core},
the spectra have been normalised to agree in the far wings of the \ion{Mg}{2}
absorption. Generally stellar \ion{Mg}{2} line profiles scale so that normalising
in the far wings produces good agreement in the inner wings for stars of the same
luminosity class \citep{2010MmSAI..81..553A}, but as we saw in Fig.~\ref{fig:mgii_core},
WASP-12's inner wings are depressed relative to other main sequence stars.
The \ion{Mg}{2} resonance line cores in WASP-12 have total absorption, which is perhaps
blue-shifted by $\sim$\,20\,\kms.

In the NUVA region of Fig.~\ref{fig:UVres_W12_alphcen}, WASP-12 exhibits similar total
absorption
in the core of the \ion{Fe}{2} resonance line at $\lambda $2586\,\AA\,
where $\alpha$\,Cen's absorption is shallower by over a factor of 10.
Appearing slightly less strong
is WASP-12's \ion{Mn}{2} resonance line at $\lambda $2577\,\AA.

%--------------------------------------------------------------------
\begin{figure}
\begin{center}
\includegraphics[width=\hsize,clip]{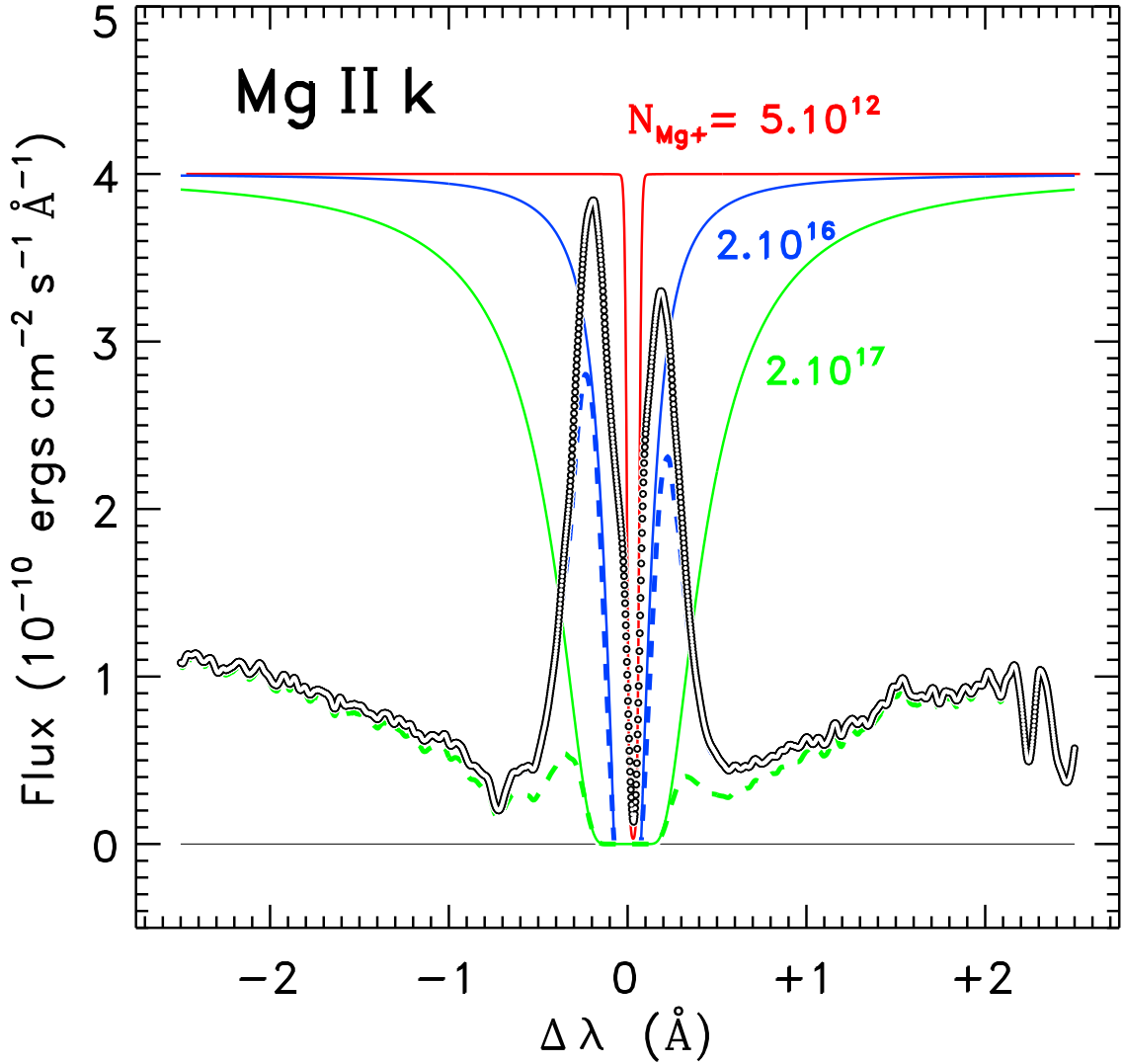}
\caption{Calculations of interstellar absorption applied to the center of the \ion{Mg}{2} line profile
of $\alpha$\,Cen. The observed line profile is plotted as open circles which overlap to
form a double line; the solid lines show the predicted absorption for column densities
of N$(Mg^{+})$ of $5\times10^{12}\, {\rm cm^{-2}}$ (red), $2\times10^{16}\, {\rm cm^{-2}}$ (blue) and
$2\times10^{17}\, {\rm cm^{-2}}$ (green). The modified line profiles produced by applying
the two higher column densities are indicated in dashed lines of corresponding colour.}
\label{fig:abs-sim}
\end{center}
\end{figure}
%--------------------------------------------------------------------

These absorption features in the WASP-12 stellar spectrum are too deep to have their origin
in the photosphere itself. Our favoured interpretation is that the absorption arises in gas within
the WASP-12 system, but we now examine the plausibility of the alternative explanation that the
absorption might arise in the ISM. Figure~\ref{fig:abs-sim}
shows the central part of the observed $\alpha$ Cen A Mg II
k-line profile. The solid red line shows the absorption we expect from the ISM along the very short
pathlength (1.3~pc) to the star,
using the Mg II column density of $5\times10^{12}$ cm$^{-2}$ reported by \citet{1996ApJ...463..254L}. The
absorption profile was
calculated using a temperature of 7000 K and a turbulent velocity of 1.5 km/s, typical values for
the local warm cloud in which $\alpha$~Cen A and the Sun are embedded. The
calculated absorption fits the narrow slightly redshifted absorption
dip in the extreme core of the line, noting that the broader ``absorption'' between the
twin peaks of the profile is a central reversal of chromospheric origin (see Linsky \& Wood
1996).  The blue curve
indicates the attenuation predicted for a
plausible N(Mg$^{+}$) in the line of sight to WASP-12: $2\times10^{16}$ cm$^{-2}$. We
arrived at this estimate assuming $E(B-V)\sim 0.2$ magnitudes for WASP-12
(based on a nominal 0.6 magnitudes kpc$^{-1}$ of reddening for an average Galactic sightline
and the $\sim 350$~pc estimated distance of WASP-12), which yields a hydrogen column density
of $\sim 1.2\times10^{21}$ from the standard Milky Way reddening law (e.g., Savage \&
Mathis 1979).
If we then
assume that roughly half the magnesium is in the form of Mg$^{+}$ --- about what is deduced from
the $\alpha$~Cen A sightline, albeit perhaps not typical --- then we obtain:
$N(Mg^{+})\sim 2\times10^{16}$ cm$^{-2}$.  Applying the corresponding absorption
profile to the $\alpha$ Cen A Mg II k line results in the dashed blue curve,
which still preserves substantial core emission.

The green line in Fig. 16 depicts the absorption profile for a Mg II column density
a factor of 10 greater than the interstellar estimate above, i.e.
$N(Mg^{+})= 2\times10^{17}$ cm$^{-2}$. Here broad damping wings have developed, strongly
suppressing
the emission core, and even affecting the inner wings of the k line.  While the predicted
profile is not an exact match to that of WASP-12 (cf., Fig. 14), it
does demonstrate the plausibility of the absorption mechanism, albeit requiring a very
substantial column of Mg$^{+}$.  The required column is excessive enough, in fact, to make an
interstellar origin seem much less plausible than a more local source, especially given
the existence of the hot Jupiter and the possibility of strong atmospheric stripping.
We examine this possibility in more detail later (\S{4.2}, below).

\section{Discussion and Conclusions}
\label{sec:conc}
The conclusions we reached in Paper 1 more or less stand in the light of our
Visit 2 data and our interpretation of the entire COS/NUV dataset. As
\S\ref{sec:res} shows, the Visit 2 data did not neatly fill in the phase-folded
NUV light curves we produced from Visit 1. Instead, a rather more complex reinterpretation
was required.
Our principle conclusions in Paper 1 (from Visit 1 data alone)
were that the NUV transit depth is deeper than the optical,
that this is due to absorption in metal atoms/ions within WASP-12\,b's exosphere, and that
this exospheric gas was spatially distributed such that an early ingress occurs.
To make sense of our combined NUV light curves, we needed to adopt two further hypotheses:
(i) the absorbing gas is (at least sometimes) even more spatially extended than we
 inferred from Visit 1; (ii) a short-lived stellar flare occurred during Visit 2 orbit 3 {\it or}
Visit 2 orbit 3 had a line of sight through a relatively low density window in the absorbing gas.

 In Section~\ref{sec:res} we presented several lines of indirect evidence in
favour of the hypothesis that
WASP-12 has some chromospheric activity, despite the resounding lack of observed
emission cores in the \ion{Mg}{2} and \ion{Ca}{2} lines. We now examine our data to see if it contains
direct evidence in favor of this hypothesis over the alternative low density window hypothesis.

\subsection{A stellar flare during Visit 2 orbit 3?}
\label{sec:flare}
\begin{figure}
\begin{center}
\includegraphics[width=\hsize,clip]{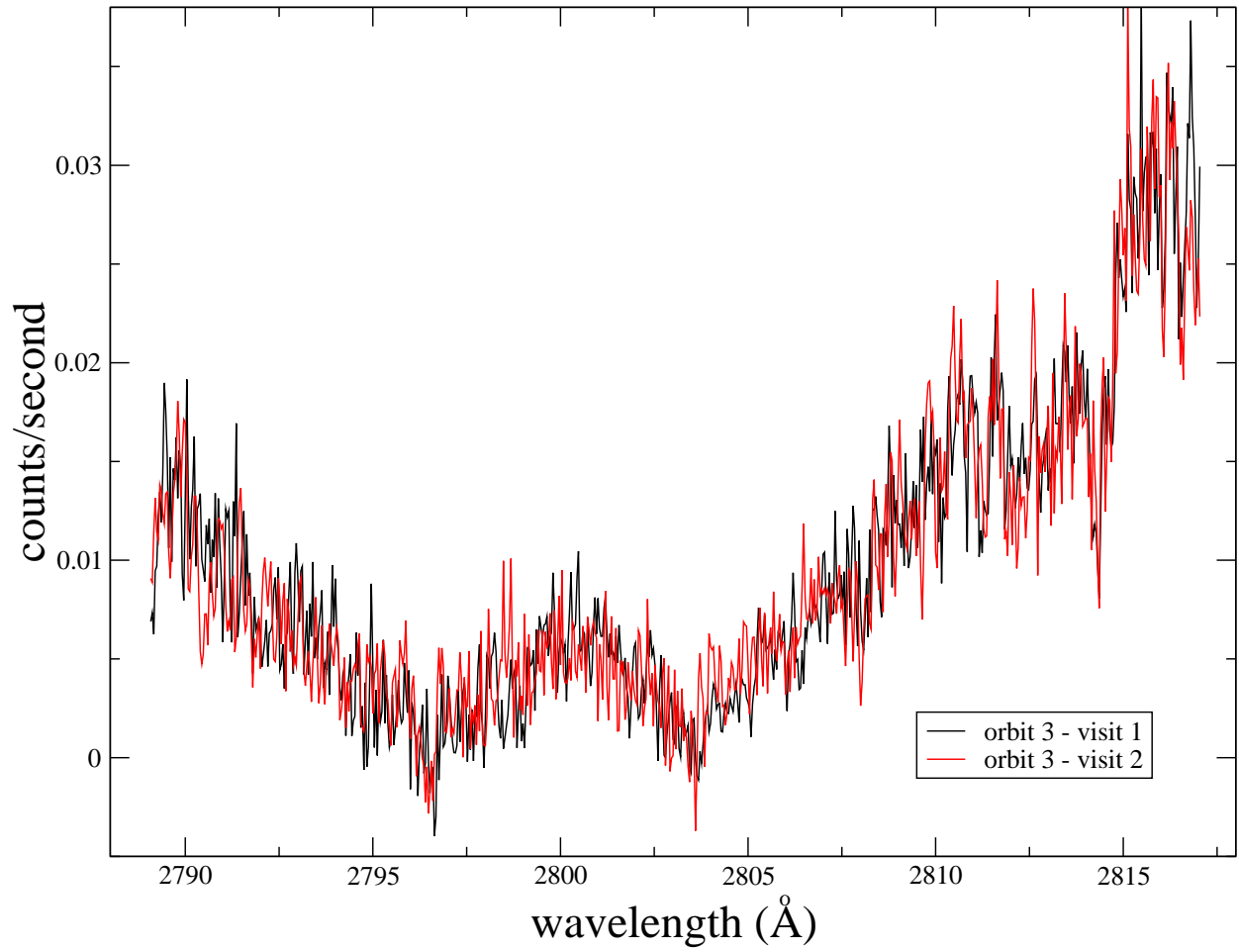}
\caption{A direct comparison of the NUVC overlapping spectral region in orbit 3 of each the two visits. }
\label{fig:comp_o3_nuvc}
\end{center}
\end{figure}

We have attributed the high NUV fluxes during Visit 2 orbit 3
to a stellar flare. To check this interpretation, Fig.~\ref{fig:comp_o3_nuvc}
shows a direct comparison of the NUV spectra from orbit 3
of Visit 2 with orbit 3 of Visit 1. The two spectra are plotted with the pipeline
calibration, no normalisation was applied to account for the declining throughput.
These spectra are clearly noisier that those shown in earlier figures
as they comprise only about a fifth of the exposure time on each visit.
The two spectra broadly agree, but there is a hint in the Visit 2 spectrum
of a small emission
component redwards of the center of each of the \ion{Mg}{2} line cores.
This is not an unassailable detection.
The appearance of these two small bumps is very similar to the remnant line
core emission on the red side of the  line profile in Figure~\ref{fig:abs-sim} after
application of absorption by a column of N${\rm (Mg^{+})} = 2\times10^{17}\, {\rm cm^{-2}}$.
The remnant emission on the blue side of the line core is absent in
Fig.~\ref{fig:comp_o3_nuvc}, but as we noted in our description of
Fig.~\ref{fig:UVres_W12_alphcen}, the absorption in WASP-12's line cores
appears slightly blue-shifted (by about $0.25 \AA$) relative to the photospheric spectral
lines. The damping wings of a blue-shifted absorption profile
(see Figure~\ref{fig:abs-sim})
would extinguish the blue side of the
emission core before the red side.
%--------------------------------------------------------------------
%--------------------------------------------------------------------

Thus Fig.~\ref{fig:comp_o3_nuvc} provides us with
(noisy) direct evidence for chromospheric emission
in the \ion{Mg}{2} line cores. Further it shows an increase in this emission {\it or} a decrease in the
absorption between the stellar chromosphere and the telescope
during Visit 2 orbit 3.

The dramatic movement
of the Visit 2 orbit 3 NUVA data point when Figs.~\ref{fig:1visfits} and \ref{fig:olap_fits}
(or Figs. ~\ref{nonorm-compv1v2} and
~\ref{nonorm-v1v2})
are compared suggests we are seeing a veiled stellar flare rather than a low density window.
In Figs.~\ref{fig:1visfits} and ~\ref{nonorm-v1v2} the \ion{Fe}{2} resonance line at $\lambda $2586\,\AA\
is included, and the flux is high. In Figs.~\ref{nonorm-compv1v2}
and ~\ref{fig:olap_fits}, this line is excluded and the flux is much lower.
A flare would increase the chromospheric emission in the \ion{Mg}{2} and \ion{Fe}{2}
resonance lines.
When the NUVA data excludes the \ion{Fe}{2} line (Fig.~\ref{fig:olap_fits}) there
is no sign of a high point in the photometry. It is only where chromospheric activity
indicators are included within the passband, i.e. always in NUVC (Figs.~\ref{fig:1visfits}, ~\ref{nonorm-compv1v2}, ~\ref{nonorm-v1v2} and ~\ref{fig:olap_fits})
and in NUVA in Figs.~\ref{fig:1visfits} and~\ref{nonorm-v1v2} but not Figs.~\ref{nonorm-compv1v2} and \ref{fig:olap_fits},
that the light curve has a high point. We infer that the enhanced chromospheric emission in \ion{Fe}{2} is absorbed by the
obscuring gas, and re-emitted isotropically and across the line profile. Thus the line core emission is
distributed across the line profile and rendered undetectable at any individual pixel
in the spectrum. When we sum over wavelength to produce light curves, however, the signal becomes apparent
above the noise.

Since the flux from the stellar photosphere in the \ion{Fe}{2} $\lambda $2586\,\AA\, line is low (Fig.~\ref{fig:UVres_W12_alphcen}), whether or not this line is included within the wavelength
coverage should make little difference to the light curve in the low density window hypothesis.
We therefore interpret the 4$\sigma$ movement of Visit 2 orbit 3 NUVA point between Fig.~\ref{nonorm-compv1v2} and Fig.~\ref{nonorm-v1v2}, i.e. between data including and excluding this line as evidence in favour
of the stellar flare hypothesis over the low density window hypothesis.
Only when we include features which we expect to show a significant flux increase
during a stellar flare do we see the NUV flux in Visit 2 orbit 3 lying significantly
above the other data.
Several aspects of our data thus suggest we caught a stellar flare
which occurred during the ingress of the optical transit in Visit 2.

\subsection{The column density of \ion{Mg}{2}}
\label{sec:column}
%--------------------------------------------------------------------
\begin{figure}
\begin{center}
\includegraphics[width=65mm]{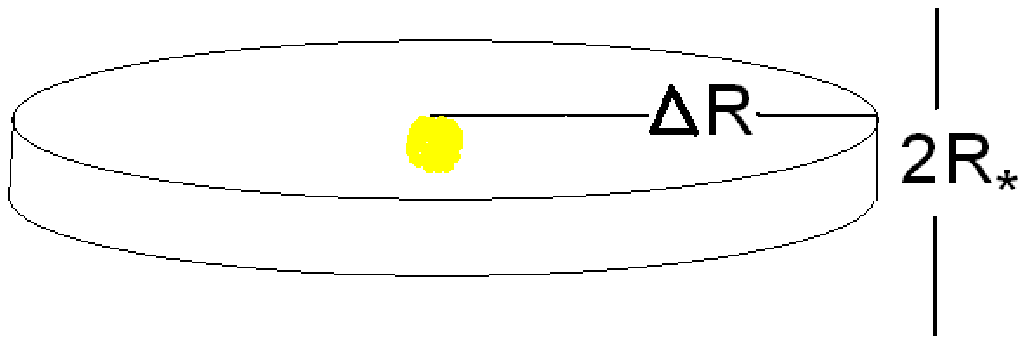}
\caption{Geometry assumed in our estimate of plausible limits to the column density due to mass loss
from WASP-12\,b. Material is assumed to be spread within a squat cylinder of radius $\Delta R$ and height
$2 R_{*}$ centered on the star.}
\label{fig:geom}
\end{center}
\end{figure}
%--------------------------------------------------------------------
In Section~\ref{sec:abs} we showed that a high column density of
\ion{Mg}{2} between us and the stellar chromosphere can absorb WASP-12's \ion{Mg}{2}
emission cores, assuming the intrinsic emission is similar to that of $\alpha$\,Cen.
Here we consider whether the value we deduced, N${\rm (Mg^{+})} \sim 2\times10^{17}\, {\rm cm^{-2}}$,
can be plausibly attributed to mass loss from the planet WASP-12\,b.
Since this extremely close-in planet is the most obviously unusual characteristic
of the star, it seems likely that the anomalous \ion{Mg}{2} line profiles are
ultimately caused by the planet.

Various mechanisms have been proposed for mass loss from WASP-12\,b and similar planets.
There are several variants of the `blow-off' hypothesis, which
assumes hydrodynamic outflow driven by energy input from irradiation by the host star.
Tidal heating can drive planetary envelope expansion and consequent Roche lobe overflow (Li et al 2010).
For the early eccentricity estimates of WASP-12\,b, for example, this mechanism was predicted to cause a mass loss rate of $10^{-7}\,{\rm M_{J}\,yr^{-1}}$ (ibid).

Mass lost from the planet will carry specific angular momentum from the planet's orbit, and will thus tend to be
confined to the orbital plane. To absorb chromospheric emission from the entire visible disc of the star,
the material must extend about one stellar radius above and below the orbital plane. To derive an upper
limit on
the absorbing column density due to mass lost from the planet, therefore, we begin with the geometry shown in Fig.~\ref{fig:geom}. Gas
is present within a squat cylinder of radius  $\Delta R$ and height $2 R_{*}$, centered on the star
and aligned with the orbital plane.

The mean density, $\bar\rho$ within the cylinder is
\begin{equation}
\bar\rho \leq \frac{\dot m \Delta t}{2 \pi \Delta R^2 R_{*}}
\end{equation}
where $\dot m$ is the mass loss rate and $\Delta t$ is the time over which
mass has been lost. The equality arises when all mass lost from the planet is confined to the cylinder we require to be populated. This is obviously an upper limit as some mass will accrete on to the star or diffuse in the vertical direction out of our line of sight to the star.

The column density of particles is
\begin{equation}
N_{tot} \leq \int_{{\rm LOS}} \frac{ \rho}{\mu} \, {\rm d}l \approx \frac{\dot m \Delta t}{2 \pi \Delta R R_{*} \mu} \approx \frac{\dot m }{2 \pi  v R_{*} \mu}
\end{equation}
where the integral is carried out along our line of sight (LOS), $\mu$ is the mean molecular weight of the gas and $v = \Delta R / \Delta t$ is the velocity with which the gas moves outwards from the orbit of the planet.
We have used the mean density and implicitly assumed that the velocity is approximately constant for outflow over the distance $\Delta R$. The integral is effectively performed over the distance $\Delta R$ which contributes
appreciably to the column density. This avoids addressing our ignorance of the functions $\rho(l)$ and $v(l)$.
The approximations
are sufficient for an order of magnitude estimate, particularly since our empirical estimate of
$v$ will ensure our approximation is weighted to the appropriate part of the velocity field.
If the fraction of the total number density in the form of \ion{Mg}{2} is $f$,
then
\begin{equation}\label{eqn:nmgplus}
N({\rm Mg^+})  \leq \frac{\dot m f}{2 \pi  v R_{*} \mu}
\end{equation}
To obtain $N({\rm Mg^+})$, we adopt the
mass loss rate predicted by
evaporation processes, for example \cite{Ehrenreich11} give
$\dot m \sim 3 \times 10^7 {\rm kg \, s^{-1}}$ for WASP-12\,b.
We take $f \approx 10^{-5}$, which is based on a number fraction of $10^{-4}$ for Mg in the solar elemental abundance mix, and a (conservative) assumption that one in 10 magnesium nuclei is in the \ion{Mg}{2} ionic state; $R_{*} = 1.1 \times 10^{9} \, {\rm m}$
\citep{Chan11}; we assume a mean molecular weight of $\mu ={\rm m_{\rm p} = 1.7 \times 10^{-27} \, {\rm kg}}$ which will be correct to within a factor of a few for material predominately composed of hydrogen.
We can estimate $v$ from our data: the absorption appears to be blue-shifted by roughly $0.25\AA$ (Fig.~\ref{fig:UVres_W12_alphcen}, 5th panel from the top), implying $v \sim 2.5 \times 10^{4} {\rm m\, s^{-1}}$
Using these values
in Eqn.~\ref{eqn:nmgplus} we obtain
$N({\rm Mg^+}) \leq 4 \times 10^{21} \, {\rm cm^{-2}}$. The required column density
is below this crude upper limit
by a factor of $\sim 2 \times 10^{4}$. Obviously, some material will move inwards and be captured by the star, and some will spread in the vertical direction
beyond the cylinder we require to be filled with gas, but we have a comfortable factor
to allow for this and other losses of \ion{Mg}{2}.

\subsection{Exospheric Gas or Coronal Gas?}
WASP-12\,b is only about one stellar diameter from the stellar surface, so this planet is orbiting
within the star's corona. This raises the possibility that absorption we observed around WASP-12b may be
attributable to entrained or bow-shocked coronal gas, rather than from mass lost from the planet \citep[e.g.][]{vidotto10,vidotto11}.
Indeed \citet{vidotto11} show how a `double' transit can arise if a transiting planet is preceded by a bow
shock in a very extended planetary magnetosphere. This does not completely
explain our Visit 2 light curves as the observed absorption begins very early and the curves which do not contain activity-indicating lines differ in shape from those which do (see \S~\ref{sec:flare}).
The typical coronal temperatures at the distance of the planet are, however, probably inconsistent with the
absorption we detect. Fig.~\ref{lam_id} gives us very strong indications that the absorption causing our NUV
transit is from gas at a similar temperature to the stellar photosphere. The densities in the gas causing the NUV
transit exceed those of a typical stellar corona (otherwise the NUV transit would be indistinguishable from the
optical transit), so naturally, by energy arguments, the temperature of this gas is lower. This argument does not,
however, rule out entrainment of coronal gas by the planet. \S~\ref{sec:column}
strongly suggests, however, that most of the gas around WASP-12 is due to mass loss from the planet: stars without
extreme planets like WASP-12\,b offer us clear sight of their \ion{Mg}{2} chromospheric emission cores.
Our observations show the emission cores are completely absorbed throughout our temporal coverage. This
requires gas at considerable distances from the planet itself.

\subsection{The Roche lobe of WASP-12\,b}
\begin{figure}
\begin{center}
\includegraphics[width=65mm]{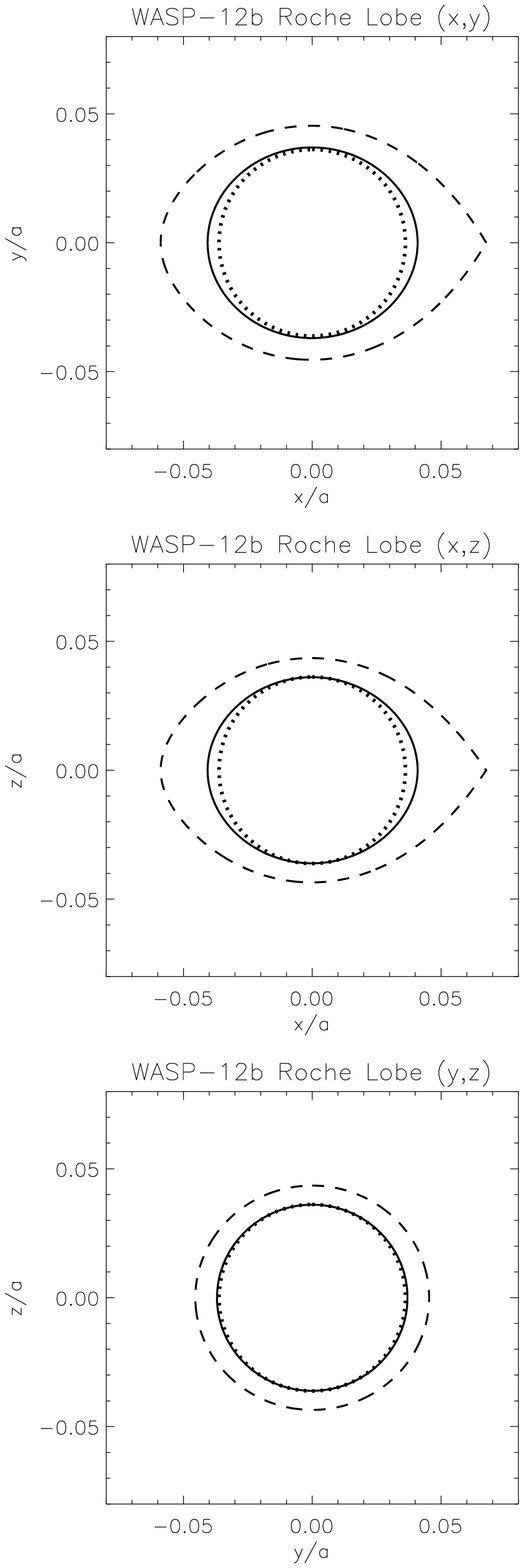}
\caption{The Roche lobe of WASP-12. The y-z cross-section (lowest panel)
is presented to us during transit. The solid black lines indicate the
equipotential surface defined by the area determined by the optical transit.
A fluid planet will fill this surface. The dotted circles have
area equal to the y-z cross-section of the planet. The
dashed lines show the critical equipotential surface which delimits the
planet's Roche lobe.}
\label{fig:roche}
\end{center}
\end{figure}

Fig.~\ref{fig:roche} shows the Roche lobe of WASP-12\,b. The y-z cross-section
is approximately that presented to us during transit; $i = 86^\circ \pm 3^\circ$ \citep{Chan11},
so the y-z plane is inclined to the
plane of the sky by $4^\circ \pm 3^\circ$.
The lowest panel of Fig.~\ref{fig:roche}
shows the cross-section of the critical equipotential surface is not much
larger than that of the planet itself. We have detected transit depths in
the NUVA and NUVC spectra regions which are up to a factor of 3 deeper than the optical
transit (see Figs.~\ref{fig:1visfits} and \ref{fig:olap_fits}). As the optical depth of the absorbing
gas will vary with both wavelength and position, it is clear that we require
the projection of the spatial distribution of the absorbing gas on the plane of the sky
to exceed that of the Roche lobe by a factor of more than a few. Our observations thus
show that the absorbing gas far overfills the planet's Roche lobe. This is consistent with
our Visit 2 data, which suggest absorption is occurring at phase 0.83 (see Fig.~\ref{nonorm-v1v2}), i.e.
the absorbing gas extends $\sim 20\%$ of the way around the orbit.
%--------------------------------------------------------------------
%--------------------------------------------------------------------

Fig.~\ref{fig:roche} also shows the deviation from spherical of WASP-12\,b. The fluid planet will fill the equipotential
surface corresponding to the area occulted in the optical transit. The y-z cross-section of this equipotential surface
deviates slightly from circular as a result of the centripetal force, with the x-y and x-z cross-sections each being
noticeably distorted by the combination of the tidal distortion due to the stellar gravity and the centripetal force.
The volume of the planet is consequently $\sim 20\%$ greater than that of a sphere producing the same optical transit depth.

\subsection{NUV versus FUV}
As we discussed in the introduction, most of the observations of diffuse gas around hot Jupiter
exoplanets have been carried out in the FUV, with our WASP-12 program being the single published exception.
Our observations are clearly limited by the SNR of the data, but it must be remembered that WASP-12
is 4 magnitudes fainter than either HD\,209458\,b or HD\,189733\,b, the two objects where FUV transit
observations exist. Even with the 4 magnitudes advantage, conclusions from the FUV transit data can be
compromised by
the fact that the FUV stellar emission is patchy and variable in time and/or severely limited by
poor SNR.
To demonstrate how FUV observations can be irrevocably compromised by these issues, we present a brief
case study using FUV COS data. The target,  HD\,189733\,b, is a hot Jupiter transiting a bright
nearby early K-star, and has been widely and informatively studied.

\begin{figure}
 \begin{center}
 \includegraphics[width=130mm,clip,angle=90]{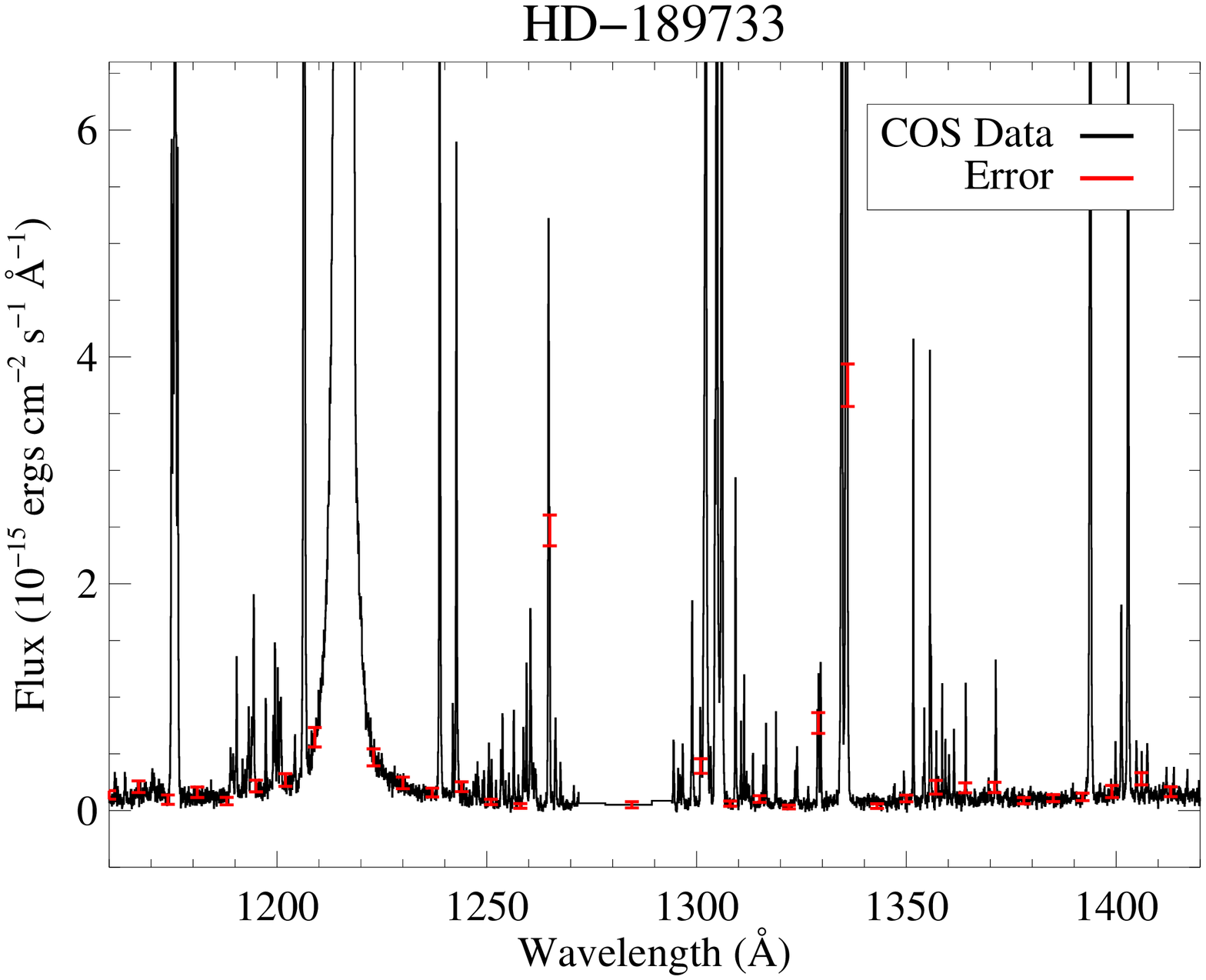}
 \caption{Coadded far-UV observations of the well-studied exoplanet host HD\,189733, obtained with
the $HST$-Cosmic Origins Spectrograph.  Despite the relatively low flux levels, COS can produce high SNR emission line spectra of planet host stars.
  }
  \label{fig:hd189733spec}
 \end{center}
 \end{figure}

The observations were obtained on 16 September 2009 as part of HST program  11673.  The data were acquired with the FUV G130M mode (Osterman et al. 2011), spanning 1134~$\leq$~$\lambda$~$\leq$~1429~\AA\ at a velocity resolution of $\approx$~17 km s $^{-1}$.
A total of 9.9 ksec of exposure time was obtained over five spacecraft orbits, allowing time for instrumental settling and stabilization~(see Vidal-Madjar et al. 2004 and references therein) prior to the transit observations.   Observations covered orbital phases from -0.074~to~0.051~(adopting the orbital parameters of Knutson et al. 2007).   The data were obtained from the Multi-Mission Archive at STScI (MAST) and calibrated one-dimensional spectra were coadded with the custom IDL procedures described by Danforth et al. (2010). The coadded spectrum is displayed in Fig.~\ref{fig:hd189733spec}. The SNR per resolution element is  $> 50$ in the
chromospheric CII 1334, 1335 A lines.
%--------------------------------------------------------------------
 %--------------------------------------------------------------------

As COS is a slitless spectrograph, geocoronal emission complicates observations at discrete wavelengths (most notably \ion{H}{1} Ly$\alpha$, 1216~\AA, and the \ion{O}{1} multiplet at 1304~\AA).  However, the high sensitivity and low instrumental background allow high-SNR observations in emission lines of several other chromospheric and transition region ions, including \ion{C}{3} $\lambda$1175, \ion{Si}{3} $\lambda$1206, \ion{N}{5} $\lambda$1240, \ion{C}{2} $\lambda$1335, and \ion{Si}{4} $\lambda$1400.
Unfortunately the very processes powering these emission lines (e.g., Ayres \& France 2010) also limit the capability of far-UV transit studies.

\begin{figure}
 \begin{center}
 \includegraphics[width=130mm,clip,angle=90]{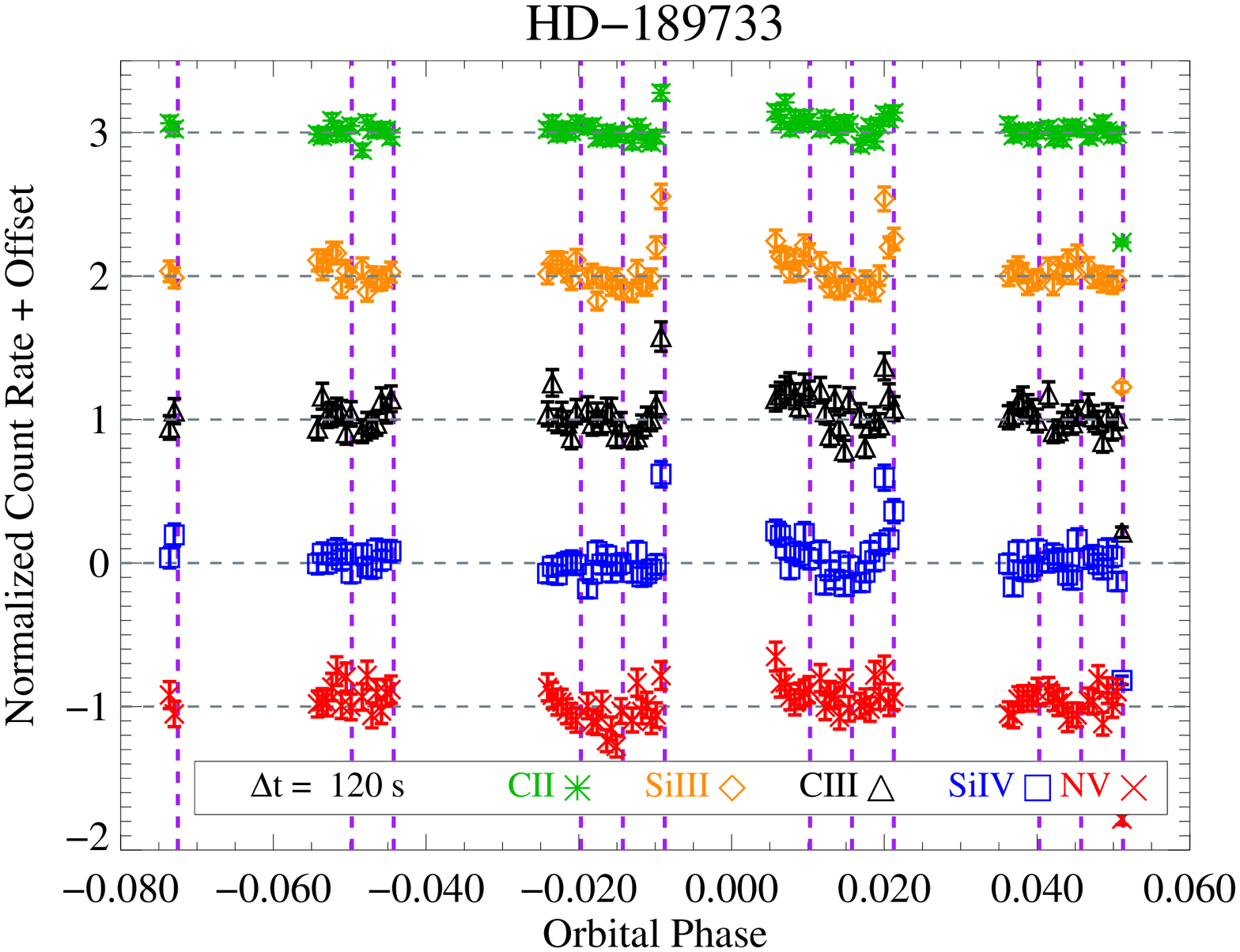}
 \caption{Count rates of the brightest far-UV emission lines not contaminated by geocoronal airglow (e.g, \ion{H}{1} Ly$\alpha$ and \ion{O}{1} $\lambda$1304).  The light curves were created from the two-dimensional COS data files, with the total number of counts in given line (defined by an [$x$,$y$] location on the COS detector) integrated in 120s intervals.  Error bars are taken to be the square-root of the number of counts in a given time bin.  We note that the variability on minute to hour timescales exceeds the photon noise and can be the limiting factor for transit studies of chromospherically active, low-mass stars. The dashed lines denote the end of an individual COS exposure.
  }
 \label{fig:hd189733lc}
 \end{center}
 \end{figure}

Light curves of the bright chromospheric emission lines from the spectrum in Fig.~\ref{fig:hd189733spec} were created from the calibrated two-dimensional data by exploiting the TIME-TAG capability (France et al. 2010). These are shown in Fig.~\ref{fig:hd189733lc}.  We extracted an [$x_{i}$,$y_{i}$,$t_{i}$] photon list from each exposure $i$ and coadded these to create a master [$x$,$y$,$t$] photon list. The total number of counts in a [$\Delta$$x$,$\Delta$$y$] box was integrated over a timestep of 120s.   The instrument background level is computed in a similar manner, with the background integrated over the same wavelength interval as the emission lines, but offset below the active science region.  The instrumental background is $<$ 6\% of the total line flux for all species presented in Fig.~\ref{fig:hd189733lc}.  Following the extraction of the light curves in photons s$^{-1}$, the data are normalized for a relative comparison of the line flux as a function of time.  For each species, we created a linear fit to the out-of-transit light curves (orbital phase $<$ -0.015 and $>$ 0.030), excluding time bins at the beginning or end of an orbit. This is similar to the normalisation procedure
 adopted in Paper 1 and in Fig~\ref{lc_norm-v1}. These normalized light curves are plotted in Fig.~\ref{fig:hd189733lc}.  The variation on timescales of minutes to hours is significantly larger than the statistical error in an individual bin.
 %--------------------------------------------------------------------
 %--------------------------------------------------------------------

If we assume that exposures taken on the fourth spacecraft orbit
represent the in-transit spectrum, we can compare the coadded,
one-dimensional in-transit and out-of-transit spectra. Integrating the total flux over airglow-free wavelength intervals, $\Delta\lambda$ = [1150~--~1210~\AA, 1220~--~1270~\AA, 1308~--~1425~\AA], we find a in-transit/out-of-transit flux ratio of 1.000~$\pm$~0.017.
These FUV observations therefore give no detection of the transit. Other FUV transit observations have given more reliable information,
 partly because HD\,209458 is an older, solar-type star with less activity than HD\,189733, but any FUV light curve results from the unknown and un-reproducible locus of the transit light curve across the patchy and changing stellar chromospheric emission distribution \citep{Haswell10}.
\citet{2012...26A...543L...4L} report STIS FUV observations of two transits of HD\,189733\,b. They too fail to detect
the transit in observations made in 2010~April. In 2011~September the STIS observations were accompanied
by contemporaneous X-ray observations in which stellar flaring activity was detected. This time the Lyman\,$\alpha$
profile showed variations attributed to exospheric absorption, but no signal
was detected in \ion{Si}{3}, \ion{O}{5} or \ion{N}{5}. \citet{2012...26A...543L...4L}, perhaps because of the brevity required for a Letter, give no assessment of the variability inherent in the observed stellar Lyman\,$\alpha$ emission.
Intriguingly, the X-ray lightcurve of \citet{2012...26A...543L...4L} shows a high point during optical ingress which is reminiscent
of our Visit 2 NUVA and NUVC orbit 3 anomaly.

Like the FUV observations of HD\,189733, our Visit 2 NUV light curve of WASP-12 was also complicated by stellar activity, but we were able to isolate and ameliorate its effect (c.f. the NUVA lightcurve in Fig.~\ref{fig:olap_fits}). The NUV can provide more robust
information on absorption by diffuse gas than the FUV. The COS/FUV observations of HD\,209458\,b \citep{linsky10} have
exquisitely high SNR;  NUV observations of this planet would clearly reveal much information about the composition
and velocity structure of its exosphere.

\subsection{Summary}
\begin{enumerate}
\item We observed WASP-12 in the NUV, a spectral region which proves to be an extremely sensitive
probe for the presence of absorbing gas.
\item We have detected asymmetric absorbing material in the WASP-12 system which is more extensive than the Roche lobe of WASP-12\,b.
\item The stellar photospheric absorption appears to be a fairly good predictor of the wavelength-dependence of the excess
absorption, indicating the extended gas may be similar in composition and ionisation state to the WASP-12 photosphere. The NUVB spectral region, which has the least photospheric absorption and is devoid of strong resonance lines yields the most straight-forward light curve. This gives us confidence in our interpretation.
\item In  HST Visit 1 we detected enhanced absorption during and before the optical transit.
\item In HST Visit 2 our NUV light curve appeared to be below the out-of-transit flux level for all the pre-ingress coverage, despite this extending to earlier phases than that of Visit 1. Thus the spatial distribution of the absorbing gas appears to vary.
\item
The cores of the \ion{Mg}{2} and \ion{Fe}{2} resonance lines are very dark throughout our
coverage. WASP-12 would be completely unique amongst all stars of similar age and
spectral type if
this lack of observed \ion{Mg}{2} chromospheric emission reversals was intrinsic to the star. We
attribute it instead to absorption occurring in gas between us and WASP-12. The
required column density is at least $\sim 2\times10^{17}$ cm$^{-2}$.  This value is
about a factor of 10
greater than a plausible estimate of the interstellar column density, but is consistent with gas
loss from WASP-12\,b via photo-evaporation.  Nevertheless, the ISM explanation cannot be dismissed
completely, since there is substantial ``cosmic variance'' in ISM properties, and the WASP-12
sightline might just be unusually opaque in Mg II.
\item The lack of \ion{Mg}{2} chromospheric emission cores from the star could be explained by  absorbing gas  not confined to the region immediately surrounding the planet. Absorbing gas would need to occupy the lines-of-sight to at least the
    mid-latitude regions of the stellar disc at all orbital phases. We show that an outflow of some fraction of the mass loss expected from WASP-12\,b can plausibly provide this obscuration, though we have not proved the gas will necessarily fulfill the geometric requirements.
\item In HST Visit 2 the NUV flux was near its maximum value at optical ingress. Such a fast change
in flux is a signature of a stellar flare, although such an event would be
very rare on a low-activity star like WASP-12.
Nevertheless, there are (albeit very noisy) hints in the \ion{Mg}{2} line profile of
some chromospheric emission at this time.
\item FUV transit observations can be far more affected by stellar activity than our NUV observations were. In the NUV we have the opportunity to isolate the effects of activity by excluding spectral lines which are strongly affected.
\item We obtained new optical photometry and combined this with archive optical photometry to derive an
improved linear ephemeris
\begin{equation}%can't work out how to re-use numbering. This is eqn 1! \eqno{1}%\tag{\ref{eq:eph}}
T_{\rm mid}({\rm HJD}) = 2454852.7739 ^{+0.00014}_{-0.00014} + N \times 1.09142206^{+0.00000033}_{-0.00000031}
\end{equation}
\end{enumerate}

\subsection{Concluding remarks}
Given its spectral type and probable age, we have suggested that
WASP-12 should have a normal level of
chromospheric activity, despite the lack of observed
emission cores in the Mg II resonance lines. This hypothesis in principle
could be tested by obtaining far-UV (FUV) spectra, which should show the
normal complement of chromospheric (e.g., O I 1305 \AA\
triplet; C II 1335~\AA\ multiplet) and higher-temperature features (e.g.,
C IV 1550~\AA\ doublet).  However, the great
distance of WASP-12 (at least 300 pc,  Fossati et al.\ 2010b) would imply line
peak flux densities below $10^{-16}$ erg cm$^{-2}$ s$^{-1} \AA^{-1}$ (scaling from the
FUV spectrum of $\alpha$~Cen A), which even with
the high sensitivity G140L grating of COS would require an exposure of $\sim 25$~ks to reach
a secure S/N$\sim 10$, or
nearly 10 {\em HST}\/ orbits.  Even worse, the estimated interstellar reddening of
$E(B-V)\sim 0.2$ magnitudes
would imply an additional factor of $\sim 5$ attenuation of the FUV lines, further
lengthening the
necessary COS or STIS exposures to beyond the realm of practicality. Furthermore,
this does not even account
for the additional absorption deduced from the WASP-12 Mg II lines, which --- if truly
interstellar and accompanied by the normal dust-to-gas ratio --- would completely
extinguish the FUV radiation.  Similarly, and again scaling from $\alpha$~Cen A, WASP-12 would
be predicted to be a 0.02 counts ks$^{-1}$ coronal X-ray source in the {\em Chandra}\/
imagers, which is about a factor of two below the average cosmic background rate.  In that
background-limited regime,
it would require at least 1~Ms of exposure to achieve a minimal 3$\,\sigma$ detection, again
completely impractical.

We need to look at more close-in exoplanets using this
highly informative NUV wavelength region. Most known hot Jupiters are less distant than WASP-12,
and consequently will offer higher NUV fluxes and thus better signal to noise.
Our conclusions throughout this paper are limited by the noise level in our data.

  Many related studies have been performed in the FUV. The FUV does provide access to higher
ionization state lines that can be good proxies for activity and suffer
less confusion with the lower-ionization ISM (CIII, NV, SiIV).
Additionally, the CIII FUV multiplet is not a resonant transition, its lower
level lies about 6.5 eV above the ground state, therefore this line is
effectively never seen in the diffuse ISM.
Despite this, as Fig. 5.10 of Haswell (2010) makes clear,
the NUV has some clear advantages over the FUV for studies of exospheric absorbing gas.
NUV transits are measured against the relatively uniform and well-understood
stellar photospheric light, whereas the FUV emission from the stellar disc is
patchy, time-variable, and limited to the wavelengths of strong chromospheric emission
lines. In addition to the consequent improvement in robustness of the inferences
drawn about the absorbing gas, the NUV also has a huge advantage in stellar flux
and thus SNR. There are many open questions and much theoretical and modeling activity
connected with the handful of extant UV
observations of transiting hot Jupiters.
It is vital that more UV transit observations are made while HST remains available.

\section*{Acknowledgments}
We gratefully acknowledge discussions with Moira Jardine, Christiane Helling and Aline
Vidotto.
This work is based on observations made with the NASA/ESA Hubble
Space Telescope, obtained from MAST at the Space Telescope Science Institute,
which is operated by the Association of Universities for Research in
Astronomy, Inc., under NASA contract NAS 5-26555. Our WASP-12 observations are
associated with program \#11651 to which support was provided by NASA through
a grant from the Space Telescope Science Institute.
This work was
supported by an STFC Rolling Grant (CAH, LF, UCK).
RB is supported by an STFC studentship.

{\it Facilities:} \facility{HST (COS)}, \facility{CFHT (ESPaDOnS)},
\facility{OHP (SOPHIE)}.

\end{document}